\RequirePackage{fix-cm}
\documentclass[final]{svjour3}       
\smartqed  
\usepackage{appendix}
\usepackage{amsmath}
\usepackage{graphicx}
\usepackage[mathlines]{lineno}
\usepackage{array}
\usepackage{longtable}
\usepackage{natbib}
\setcitestyle{aysep={}}

\usepackage[hidelinks]{hyperref}
\usepackage{amssymb}
\usepackage{latexsym}
\usepackage{xcolor}

\newcommand*\patchAmsMathEnvironmentForLineno[1]{%
\expandafter\let\csname old#1\expandafter\endcsname\csname #1\endcsname
\expandafter\let\csname oldend#1\expandafter\endcsname\csname end#1\endcsname
\renewenvironment{#1}%
{\linenomath\csname old#1\endcsname}%
{\csname oldend#1\endcsname\endlinenomath}}%
\newcommand*\patchBothAmsMathEnvironmentsForLineno[1]{%
\patchAmsMathEnvironmentForLineno{#1}%
\patchAmsMathEnvironmentForLineno{#1*}}%
\AtBeginDocument{%
\patchBothAmsMathEnvironmentsForLineno{equation}%
\patchBothAmsMathEnvironmentsForLineno{align}%
\patchBothAmsMathEnvironmentsForLineno{flalign}%
\patchBothAmsMathEnvironmentsForLineno{alignat}%
\patchBothAmsMathEnvironmentsForLineno{gather}%
\patchBothAmsMathEnvironmentsForLineno{multline}%
}

\begin{document}

\title{Turbulence organization and mean profile shapes in the stably stratified boundary layer: zones of uniform momentum and air temperature}

\titlerunning{Turbulence organization in the stably stratified boundary layer}

\author{Michael Heisel \and Peter P. Sullivan \and Gabriel G. Katul \and Marcelo Chamecki}

\institute{M. Heisel \at
              Department of Atmospheric \& Oceanic Sciences, University of California Los Angeles, Los Angeles, CA 90095, USA\\
              \email{heisel@ucla.edu}           
           \and
           P. P. Sullivan \at
              National Center for Atmospheric Research, Boulder, CO 80301, USA
           \and
           G. G. Katul \at
              Department of Civil and Environmental Engineering and Nicholas School of the Environment, Duke University, Durham, NC 27708, USA
           \and
           M. Chamecki \at
           Department of Atmospheric \& Oceanic Sciences, University of California Los Angeles, Los Angeles, CA 90095, USA
}

\date{Submitted: DD Month YEAR}

\maketitle

\begin{abstract}

A persistent spatial organization of eddies is identified in the lowest portion of the stably-stratified planetary boundary layer. The analysis uses flow realizations from published large-eddy simulations (Sullivan et al., J Atmos Sci 73(4):1815–1840, 2016) ranging in stability from neutral to nearly z-less stratification. The coherent turbulent structure is well approximated as a series of uniform momentum zones (UMZs) and uniform temperature zones (UTZs) separated by thin layers of intense gradients that are significantly greater than the mean. This pattern yields stairstep-like instantaneous flow profiles whose shape is distinct from the mean profiles that emerge from long-term averaging. However, the scaling of the stairstep organization is closely related to the resulting mean profiles. The differences in velocity and temperature across the thin gradient layers remain proportional to the surface momentum and heat flux conditions regardless of stratification. The vertical thickness of UMZs and UTZs is proportional to height above the surface for neutral and weak stratification, but becomes thinner and less dependent on height as the stability increases. Deviations from the logarithmic mean profiles for velocity and temperature observed under neutral conditions are therefore predominately due to the reduction in zone size with increasing stratification, which is empirically captured by existing Monin-Obukhov similarity relations for momentum and heat. The zone properties are additionally used to explain trends in the turbulent Prandtl number, thus providing a connection between the eddy organization, mean profiles, and turbulent diffusivity in stably stratified conditions.

\end{abstract}

\section{Introduction}
\label{sec:intro}

Since the introduction of Monin-Obukhov (M-O) similarity theory \citep{Monin1954,Foken2006}, significant effort has been devoted to fine-tuning the empirical similarity relations $\phi(z/L)$ \citep[e.g.,][]{Businger1971,Dyer1974,Hogstrom1988}. A logarithmic wind profile is expected for neutrally-stratified conditions in the planetary boundary layer (PBL), and $\phi$ accounts for deviations from the logarithmic profile due to thermal stability effects. The basis of the theory is that $\phi$ is a universal function of $z/L$ for a given dimensionless quantity such as the mean shear, where $z$ is height above the surface and $L$ is the Obukhov length defined using surface scaling parameters \citep{Obukhov1946}. Due to limitations in measuring and simulating atmospheric turbulence, the fine-tuning efforts have not been accompanied by a concrete understanding of how the organization and intensity of turbulent eddies quantitatively changes with varying stratification. Aspects of the eddies can be inferred from various techniques including visualization by fog and clouds \citep{Young2002}, spectra of point measurements \citep{Kaimal1972}, refractive indexing \citep{Wyngaard2001}, correlations across an array of sensors \citep{Salesky2013,Lan2018}, and integral scales from simulations \citep{Huang2013,Chinita2022}, but there is scant evidence supporting a formal link between the eddy properties and $\phi(z/L)$ \citep{Katul2011}.

The prevailing organization of turbulent motions depends on the stability regime. These motions have been widely observed in the convective PBL due to the large and long-lived scale of the dominant eddies; the prominence of elongated streaks near the surface in neutral conditions \citep{Young2002}, roll vortices of various sizes in weak convection \citep{Etling1993}, and open cells dominated by buoyancy in forced convection \citep{Atkinson1996} have all been observed experimentally and with large-eddy simulations (LES) \citep{Khanna1998,Shah2014,Patton2016,Salesky2017}. Relatively less is known about the organization of turbulent eddies in stably-stratified conditions \citep[see][for a review on the topic]{Mahrt2014}. Average flow statistics reveal distinct scaling regimes for the stable PBL \citep{Holtslag1986,Mahrt1999}: in the near-surface region under weak stratification, the traditional M-O surface parameterization appears applicable; farther from the surface, similarity is better represented using local, height-dependent fluxes in the definition of $L$ \citep{Nieuwstadt1984,Sorbjan1986}; under increased stability, the flow approaches z-less stratification for which $z$ is no longer a relevant parameter \citep{Wyngaard1972}; finally, very stable conditions yield large-scale global intermittency throughout the PBL depth \citep{Mahrt1989}. Simulations for weak stratification at relatively low Reynolds number found turbulent features similar to neutral conditions \citep{Garcia2011,Watanabe2019,Atoufi2021}. It is reasonable to expect the properties of these eddies to be consistent with scaling of the more stable regimes, e.g. a loss of height dependence in eddy size for z-less stratification, but a detailed assessment relating the turbulent eddy organization and average flow statistics in stably-stratified conditions has not been made to date and frames the scope here.

Recent advances in turbulent boundary layer research can be used as a starting point for such an assessment. Instantaneous realizations of neutrally-stratified wall-bounded flows can be approximated as a population of large regions with relatively uniform along-wind momentum. These regions are separated by thin layers characterized by enhanced shear and vorticity \citep{Meinhart1995,Priyadarshana2007}. The large regions are known as uniform momentum zones (UMZs) \citep{Meinhart1995,deSilva2016} and the thin layers associated with the UMZ edges have various descriptors including internal shear layers \citep{Gul2020}, internal interfacial layers \citep{Fan2019}, and vortical fissures \citep{Priyadarshana2007}. Geometric facets within the UMZs exhibit self-similarity that has been associated with statistics in the inertial subrange of scales \citep{Heisel2022}. Further, the intermittent spacing of small-scale features concentrated within the thin layers has been identified for more general flow conditions \citep[e.g.,][]{Ishihara2009} including direct observation in the atmospheric surface layer \citep{Heisel2018}, such that this observed eddy organization is considered a potential archetypal turbulent structure \citep{Hunt2010,Elsinga2010,Ishihara2013,Hunt2014}.

The UMZs and shear layers are not independent of previously identified features because the generic definition of UMZs applies to any coherent velocity region. For example, the signature of UMZs in the inertial layer correspond closely to low-momentum streaks that are often sandwiched by high-momentum counterparts \citep{Hutchins2007,Dennis2011,Smits2011}. These streaks have been observed for atmospheric flows using arrays of point sensors \citep{Wilczak1980,Hutchins2012}, doppler lidar \citep{Traumner2015}, and large-eddy simulations \citep{Salesky2018}. The low-momentum streaks -- and the signature of a UMZ -- have been associated with packets of hairpin-like vortices aligned with the high-shear region along the upper edge of the UMZ \citep{Adrian2000,Adrian2007}. Similar vortical features have been identified in measurements \citep{Hommema2003,Carper2004,Hutchins2012,Heisel2018} and LES \citep{Lin1996} of atmospheric flows. A key advantage of detecting UMZs and their edges as opposed to streaks and hairpin-like vortices is that it allows for a quantitative link between eddy organization and mean flow statistics. For instance, in neutral conditions the characteristic velocity of the shear layers is the friction velocity $u_*$ and the vertical thickness of UMZs is proportional to $z$ \citep{Heisel2020}. These parameters match the behavior assumed in derivations of the logarithmic (log) law such as Prandtl's mixing length closure \citep{Prandtl1932} and Townsend's attached eddy hypothesis \citep{Townsend1976}, thus identifying the organization of eddies in physical space underlying the logarithmic wind profile that forms the foundation of M-O similarity. Simplified models based on these eddy properties can reproduce first- and higher-order flow statistics for canonical boundary layers \citep{deSilva2017,Marusic2019,Bautista2019}.

The same organization of eddies has been proposed for the turbulent temperature field in stratified conditions, namely uniform temperature zones (UTZs) and thin layers with concentrated temperature gradients \citep{Ebadi2020}. These thin gradient layers have been observed in simulations of stably-stratified shear turbulence \citep{Chung2012,Glazunov2019} and are traditionally known as fronts \citep{Chen1978,Sullivan2016} or microfronts \citep{Mahrt1994,Mahrt2019} separating regions of differing temperature. Similar to the thin shear layers, hairpin-like vortices have been detected along the fronts \citep{Mahrt1994,Sullivan2016}. It is likely that the fronts are related to the cliffs of the ramp-cliff pattern common in time series of temperature and other scalars \citep{Antonia1979,Kikuchi1985,Warhaft2000}. Further, the persistent presence of the fronts and ramp-cliff structure yields a stairstep-like shape in instantaneous profiles of the temperature \citep{Sullivan2016}. The layering of the turbulent eddies and the corresponding stairstep profile have been observed for stratified flows in applications beyond the atmosphere \citep[see, e.g.,][and references therein]{Praud2005,Basak2006,Waite2011,Caulfield2021}. These patterns are analogous to the stairstep profile of UMZs and shear layers seen for velocity \citep{deSilva2016,Heisel2020}. Hence, the UMZ framework can similarly be applied to the temperature field to quantify the vertical thickness of UTZs (i.e. vertical spacing between fronts) and the intensity of the temperature difference across the front to relate the turbulent temperature structure to the mean profile.

The present work uses existing techniques for detecting UMZs and UTZs to investigate the organization of turbulent eddies in the stably-stratified PBL. Turbulent flow volumes are analyzed from a previously-published suite of high-resolution LES \citep{Sullivan2016}. The LES conditions range from neutrally-stratified to near z-less stratification and exclude the very stable regime with global intermittency. The region of interest is the lower portion of the PBL including heights within and closely above the surface layer where a transition from surface to local to z-less scaling is expected. The study seeks to address two key questions motivated by the above introduction: (i) is the organization of turbulent eddies in the stably-stratified PBL qualitatively similar to neutral conditions (i.e. a series of UMZs and UTZs)? (ii) how do changes in the eddy properties with increasing stratification relate to log law mean profile deviations that are predicted empirically by $\phi(z/L)$? The remainder of the article is organized as follows: the LES and detection of uniform zones are detailed in Sec. \ref{sec:methods}; properties of the detected zones are presented in Sec. \ref{sec:results}; concluding remarks are given in Sec. \ref{sec:conclusion}.

\section{Methodology}
\label{sec:methods}

The first four subsections provide a detailed account for each aspect of the analysis: simulation of the PBL (\ref{subsec:les}), rotation of the numerical grid in post-processing (\ref{subsec:orientation}), detection of UMZs (\ref{subsec:detection}), and detection of UTZs (\ref{subsec:detection_utz}). An abbreviated summary of the methodology is given in Sec. \ref{subsec:summary} for readers only interested in the details essential for understanding the results in Sec. \ref{sec:results}.

\subsection{Planetary boundary layer simulations}
\label{subsec:les}

The analysis is conducted on previously-published LES of the stably-stratified PBL \citep{Sullivan2016}. A new simulation for the neutrally-stratified PBL is also included here to assess the transition from neutral to stable conditions. A complete description of the stratified simulations \citep{Sullivan2016} and similar neutral simulations \citep{Moeng1994,Lin1996} are provided elsewhere, and an overview of relevant details is given here.

The LES design follows the GEWEX Atmospheric Boundary Layer Study (GABLS) benchmark case for a canonical stable PBL \citep[see, e.g.,][]{Beare2006,Basu2006,Cuxart2006,Huang2013,Matheou2014}. The geostrophic wind ($U_g=$ 8 m\,s$^{-1}$), high-latitude Coriolis frequency ($f=$ 1.39 $\times$ 10$^{-4}$ s$^{-1}$), still air initial potential temperature ($\theta_o=$265 K), capping inversion strength (0.01 K\, m$^{-1}$), domain size (400 m in each direction), and fixed surface cooling rate ($C_r=$ 0.25 K\,h$^{-1}$) of the benchmark case were all adopted for the present LES. Three additional higher cooling rates up to $C_r=$ 1 K\,h$^{-1}$ were also simulated. The neutral PBL was modeled using the same flow conditions in the absence of surface cooling.

The stably-stratified LES was computed on a 1024$^3$ numerical grid, yielding an isotropic resolution $\Delta=$ 0.39 m in each direction. The neutral conditions were discretized on a relatively smaller 512$^3$ grid corresponding to $\Delta=$ 0.78 m. Consideration of the difference in resolution is discussed in Sec. \ref{subsec:detection}. For all cases, the LES employed a two-part subgrid-scale (SGS) model detailed elsewhere \citep{Sullivan1994}. The surface conditions at the bottom boundary were estimated using an M-O similarity wall model \citep{Moeng1984} applied locally to each point along the horizontal surface \citep{Mironov2016}.

The scaling parameters resulting from each simulated surface cooling rate are listed in Table \ref{tab1}. Here, $Q_* = \langle w^\prime \theta^\prime \rangle (z{=}0)$ is the surface heat flux, $u_* = \sqrt{\langle u^\prime w^\prime \rangle_h (z{=}0)}$ is the friction velocity analogous to the surface momentum flux, $\theta_* = - Q_*/u_*$ is the surface temperature scaling, and the Obukhov length definition $L= - u_*^2 \theta_o / \kappa g \theta_*$ is based on the von K\'{a}rm\'{a}n constant $\kappa \approx 0.4$ and gravitational constant $g$. Angled brackets ``$\langle \cdot \rangle$'' imply averaging across both time and the horizontal plane, prime notations ($^\prime$) indicate fluctuations around the average value, e.g. $u^\prime = u-\langle U \rangle$, and the subscript $h$ is shorthand for the magnitude of the horizontal velocity ($u$ and $v$) contributions, e.g. $\langle u^\prime w^\prime \rangle_h = \sqrt{\langle u^\prime w^\prime \rangle^2 + \langle v^\prime w^\prime \rangle^2 }$. The virtual potential temperature $\theta$ is hereon simplified as ``temperature''.

\begin{table}
\caption{Key scaling parameters for large-eddy simulations (LES) of the neutral and stably-stratified planetary boundary layer. A full account of the stable LES is given elsewhere \citep{Sullivan2016}.}
\label{tab1}       
\begin{tabular}{cllllllll}
\hline\noalign{\smallskip}
Symbol															& $C_r$			& $Q_* \times 10^3$	& $u_*$  		& $\theta_*$	& L			& $z_i$	& $z_i/L$	& $h/z_i$	\\
																& (K\,h$^{-1}$)	& (K\,m\,s$^{-1}$)	& (m\,s$^{-1}$)	& (K)			& (m)		& (m)	& (--)		& (--)		\\
\noalign{\smallskip}\hline\noalign{\smallskip}
$\boldsymbol{\bigtriangledown}$									& 0				& 0					& 0.332			& 0				& $\infty$	& 263	& 0			& 0.98		\\			
\textcolor[rgb]{0.2471,0.0353,0.5655}{$\boldsymbol{\bigcirc}$}	& 0.25			& --9.63			& 0.255			& 0.0378		& 116		& 198	& 1.7		& 0.81		\\
\textcolor[rgb]{0.6941,0,0.5208}{$\boldsymbol{\times}$}			& 0.375			& --11.5			& 0.234			& 0.0493		& 74.7		& 182	& 2.4		& 0.75		\\
\textcolor[rgb]{0.9294,0.3788,0.2259}{$\boldsymbol{\triangle}$}	& 0.5			& --13.5			& 0.222			& 0.0607		& 54.7		& 173	& 3.2		& 0.71		\\
\textcolor[rgb]{0.9325,0.7451,0.2094}{$\boldsymbol{+}$}			& 1				& --19.5			& 0.194			& 0.100			& 25.5		& 154	& 6.0		& 0.58		\\
\noalign{\smallskip}\hline
\end{tabular}
\end{table}

The depth of the PBL is parameterized here as the inversion height $z_i$ based on the position where the temperature gradient $\partial \langle \theta \rangle / \partial z$ is maximized \citep{Sullivan1998}. Another common definition for the PBL depth in stable conditions is the height where the average shear stress is a small fraction of the surface value. This depth $h = z(\langle u^\prime w^\prime \rangle_h{=}0.05u_*^2)/0.95$ \citep{Kosovic2000} is closely related to the position of the low-level jet \citep{Blackadar1957}. For the neutral LES case, $L \sim O(10^5)$ is considered infinite and buoyancy effects below the inversion layer are considered negligible for the purposes of the study. The stable cases (in order of increasing $C_r$) correspond to runs C, D, E, and F in the original study \citep{Sullivan2016}.

\begin{figure}
\centering
  \includegraphics[width=\linewidth]{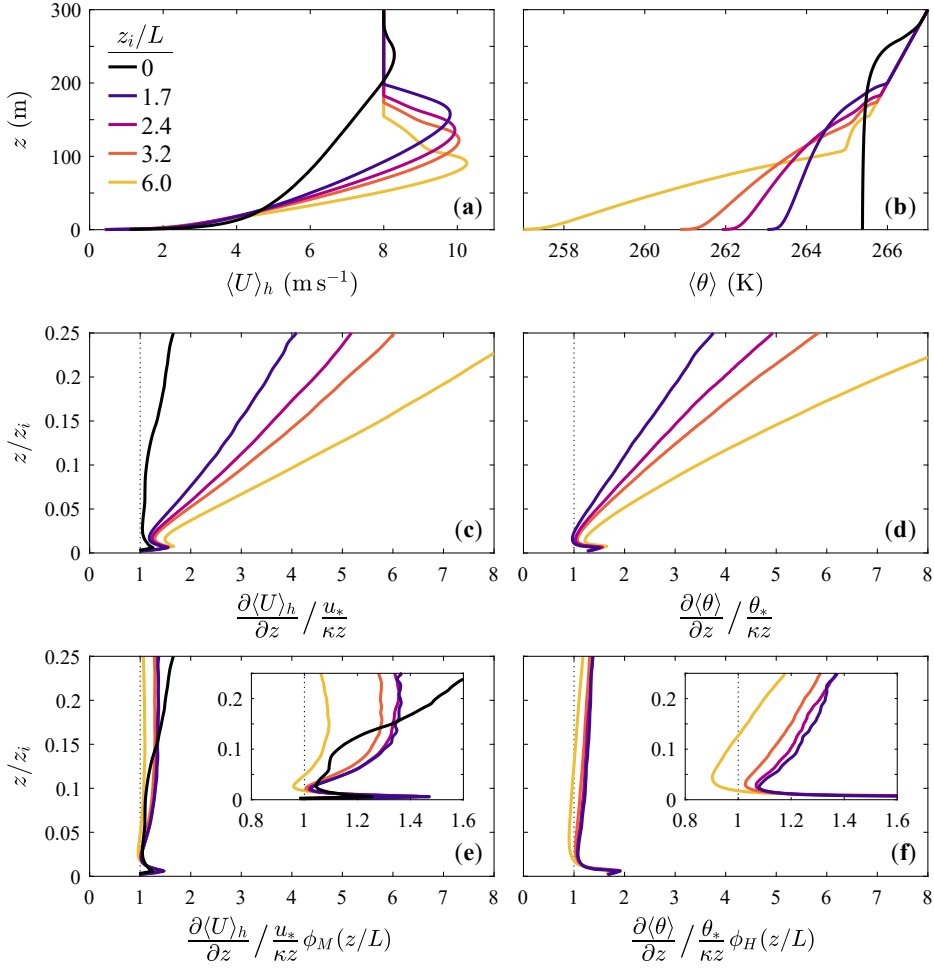}
\caption{Mean profiles for the large-eddy simulations (LES) of a stably-stratified planetary boundary layer: \textbf{a} horizontal wind speed magnitude $\langle U \rangle_h = \sqrt{\langle U \rangle^2 + \langle V \rangle ^2 }$; \textbf{b} virtual potential temperature $\langle \theta \rangle $; \textbf{c} velocity gradient; \textbf{d} temperature gradient; \textbf{e} velocity gradient with the similarity relation $\phi_M = 1+4.7 z/L$ for momentum; \textbf{f} temperature gradient with the similarity relation $\phi_H = 0.74+4.7 z/L$ for heat. In this and later figures, the legend corresponds to the $z_i/L$ stability parameter for each LES case in Table \ref{tab1}.}
\label{fig:mean_profiles}     
\end{figure}

Profiles of the mean horizontal wind speed and air temperature are shown in Figs. \ref{fig:mean_profiles}a and \ref{fig:mean_profiles}b for each LES case in Table \ref{tab1}. Increased surface cooling leads to stronger mean gradients in $\langle U \rangle_h$ and $\langle \theta \rangle$, an increased super-geostrophic maximum jet speed, and a reduction in the PBL depth. The trend most relevant to the present analysis is the behavior of the mean gradients near the surface and away from the influence of the jet. Accordingly, the bottom 25\% of the PBL based on $z_i$ is chosen as the region of interest. In the context of the turbulent kinetic energy (TKE) budget, there is an approximate local equilibrium between shear production, buoyancy damping, and dissipation of energy at each height within this region. This range includes heights above the surface layer, i.e. beyond 0.1$h$, where the approach to z-less stratification is more evident.

The mean gradients are compared to the log law scaling parameters in Figs. \ref{fig:mean_profiles}c and \ref{fig:mean_profiles}d, where $\partial \langle U \rangle_h / \partial z=u_*/\kappa z$ is expected within the surface layer for neutral conditions, yielding a logarithmic dependence on $z$ upon integration. As predicted by M-O similarity, the deviation from log law scaling increases with both height $z$ and stratification $L^{-1}$. Figures \ref{fig:mean_profiles}e and \ref{fig:mean_profiles}f include corrections to the gradients $\partial \langle U \rangle_h / \partial z=(u_*/\kappa z) \phi_M$ and $\partial \langle \theta \rangle / \partial z=(\theta_*/\kappa z) \phi_H$ based on common empirical similarity relations for momentum $\phi_M = (1+4.7z/L)$ and heat $\phi_H = (0.74+4.7z/L)$ \citep{Businger1971}. The similarity relations account for a large majority of the deviation from log law scaling of the mean profiles in the lowest 25\% of the PBL. Yet, the residual differences are non-negligible and exhibit a possible stability trend, as evidenced by the inset panels of Figs. \ref{fig:mean_profiles}e and \ref{fig:mean_profiles}f. However, the focus of this study is the large deviations from log law scaling and how these deviations are related to the underlying eddy organization. The following sections detail the detection of these eddies and assess the eddy properties in the context of the mean profiles.

\subsection{Grid orientation}
\label{subsec:orientation}

\begin{figure}
\centering
  \includegraphics[width=\linewidth]{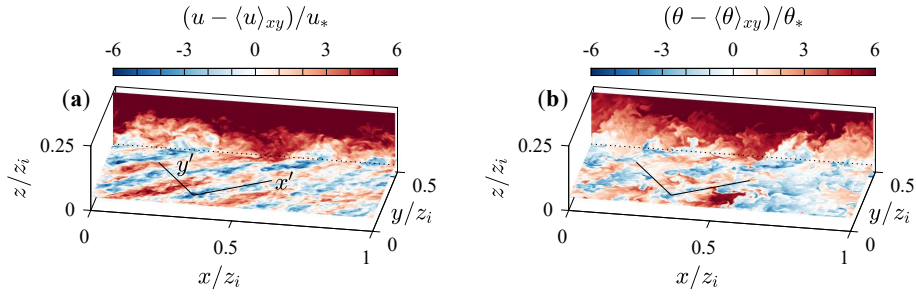}
\caption{Example regions of coherent momentum (\textbf{a}) and temperature (\textbf{b}) visualized along the $x{-}y$ and $x{-}z$ planes. The example is from the weakly stable case with $z_i/L = 1.7$. The $x{-}y$ plane shown corresponds to $z=0.05z_i$. The ($x^\prime$, $y^\prime$) notation refers to the horizontal coordinates oriented with the mean wind direction at the surface.}
\label{fig:streaks}     
\end{figure}

For neutral and weak stratification, the coherent velocity regions in the inertial layer organize as elongated streaks of low and high momentum \citep[e.g.][]{Hutchins2007,Garcia2011}. These streaks are apparent in the $x{-}y$ horizontal plane of instantaneous flow realizations such as the example in Fig. \ref{fig:streaks}a. Here, ``$\langle \cdot \rangle_{xy}$'' indicates the spatial average at the height of the horizontal plane $z=0.05z_i$. Similar features are present in the Fig. \ref{fig:streaks}b turbulent temperature field, but there is notably less coherence across longer distances such that the uniform temperature regions do not appear as ``streaks''. There is significant overlap in the regions of negative and positive fluctuations in \ref{fig:streaks}a and \ref{fig:streaks}b which has been similarly observed for convective flows \citep{Khanna1998,Krug2020}.

The coherent velocity and temperature regions cross through the $x{-}z$ vertical plane that is also shown in Fig. \ref{fig:streaks}, leaving a visible signature in this plane. However, due to Coriolis effects the near-surface turbulent features are not oriented with the geostrophic wind direction ($x$). To align the forthcoming analysis with the orientation of these features, the horizontal plane was rotated according to the mean wind direction in the lowest 10\% of the PBL. The rotated coordinates ($x^\prime$,$y^\prime$) are indicated in Fig. \ref{fig:streaks} and are closely aligned with the velocity streaks. Compared to the original $x{-}y$ plane, the uniform velocity and temperature regions are apparent across larger distances along $x^\prime$, which greatly facilitates the detection of the zones. The velocity component along $x^\prime$, given the notation $u_{x^\prime}$, is calculated by first rotating the horizontal velocity along the new ($x^\prime$,$y^\prime$) coordinate system at the original grid points, then using two-dimensional linear interpolation to estimate the velocities at new grid points aligned with $x^\prime$. The interpolation employed the same grid spacing $\Delta$ as the simulations and was conducted \textit{a posteriori} on output flow volumes.

\begin{figure}
\centering
  \includegraphics[width=\linewidth]{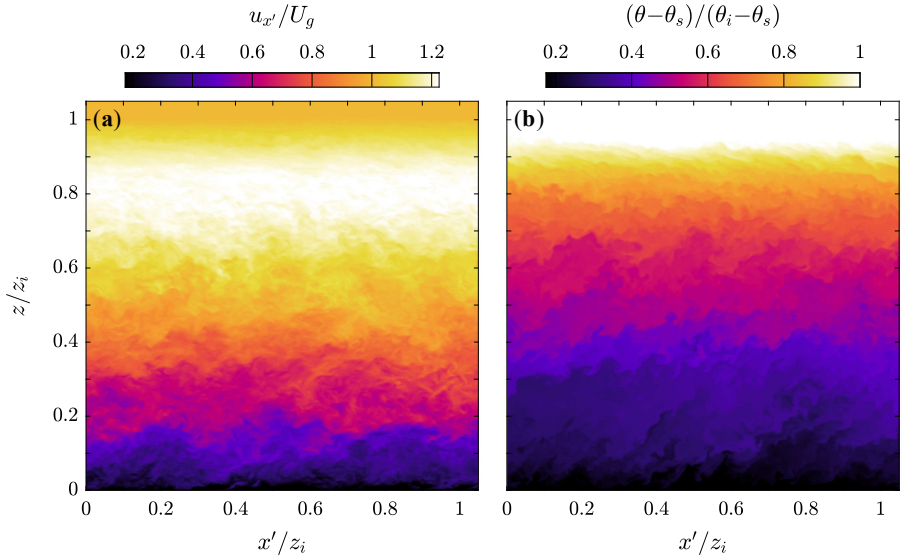}
\caption{Example velocity (\textbf{a}) and temperature (\textbf{b}) fields along the $x^\prime{-}z$ plane oriented with the mean wind direction at the surface. The example is from the weakly stable case with $z_i/L = 1.7$. The values are normalized by the geostrophic wind speed $U_g$, mean surface temperature $\theta_s$, and mean temperature $\theta_i$ at the inversion height $z_i$.}
\label{fig:fields}     
\end{figure}

Figure \ref{fig:fields} shows example instantaneous realizations of $u_{x^\prime}$ and $\theta$ interpolated along the $x^\prime{-}z$ plane. The fields are normalized such that the values range from 0 at the surface to 1 at $z_i$, with the notable exception of the super-geostrophic wind speed for the low-level jet. The organization of the instantaneous turbulence into relatively uniform zones is apparent for both quantities as indicated by similarity in color across large regions in the lower half of the PBL. The temperature field in Fig. \ref{fig:fields}b appears to be more well-mixed within the zones and exhibits more distinct temperature fronts visible from abrupt changes in color. The thin shear layers are less striking in Fig. \ref{fig:fields}a, but large velocity differences are apparent at multiple heights including $z\approx$ 0.15$z_i$ and 0.3$z_i$.

The LES resolution is likely a critical factor for the simulations to reproduce the observed uniform zones. Coarser resolution will yield a larger effective eddy viscosity such that the SGS dynamics along the gradient layers will be spread across wider distances, i.e. the gradient layers will appear less ``thin''. If $\Delta$ is comparable to the expected zone size, the distinction between the zones and the gradient layers along their edges will be lost. The same is true for experiments at low Reynolds number, where sufficient scale separation between the inertia-dominated zones and the viscous shear layers is required for the UMZ organization to become apparent \citep[see, e.g.,][]{deSilva2017}. The grid resolution also influences the statistical detection of zones as discussed in the following section.

\subsection{Uniform momentum zone detection}
\label{subsec:detection}

Because the along-wind velocity component within a UMZ is relatively uniform by definition, the velocities at grid points within the zone tend to cluster around a single value representative of the UMZ. This tendency yields a statistical signature for the UMZs, where the grid points with similar velocity value manifest as a distinct peak in velocity histograms of instantaneous flow realizations \citep{Adrian2000,deSilva2016}. The automated detection of UMZs (and UTZs) from these histograms is sensitive to design parameters for both computing the histogram and identifying its peaks \citep{deSilva2016,Laskari2018,Heisel2018}. These sensitivities make it necessary to correctly fix the parameters for all flow cases within a given study, and make it challenging to conduct quantitative comparisons across different studies that employ varying detection parameters. The conclusions drawn here thus focus on robust trends across the flow cases rather than exact quantitative properties of the detected zones.

The foremost design parameter is the size of the local flow volume used to compute each histogram, where the size is often expressed in terms of the along-wind distance $\mathcal{L}_{x^\prime}$. The appropriate scaling for $\mathcal{L}_{x^\prime}$ depends on the specific study, where the consequences for various scaling options are discussed elsewhere \citep{deSilva2016,Heisel2018,Heisel2020}. If $\mathcal{L}_{x^\prime}$ is large relative to the extent of UMZs near the surface, numerous UMZs and their velocity clusters blend together such that the distinct histogram peaks are lost. In contrast, a small $\mathcal{L}_{x^\prime}$ can yield spurious peaks due to measurement noise and resolution limitations \citep[see, e.g. Fig. 8 of][]{Heisel2018}. The distance $\mathcal{L}_{x^\prime}=0.1z_i$ employed here matches the value of a similar study that evaluated properties of UMZs in the inertial layer \citep{Heisel2020}. A sensitivity analysis demonstrated that increasing $\mathcal{L}_{x^\prime}$ leads to the detection of fewer UMZs, but the agreement across flow cases does not change if $\mathcal{L}_{x^\prime}/z_i$ is matched \citep{Heisel2020}. In other words, fixing $\mathcal{L}_{x^\prime}$ relative to a fraction of the PBL depth ensures the effects of UMZ blending and histogram convergence are the same across each case in Table \ref{tab1}.

\begin{figure}
\centering
  \includegraphics[width=\linewidth]{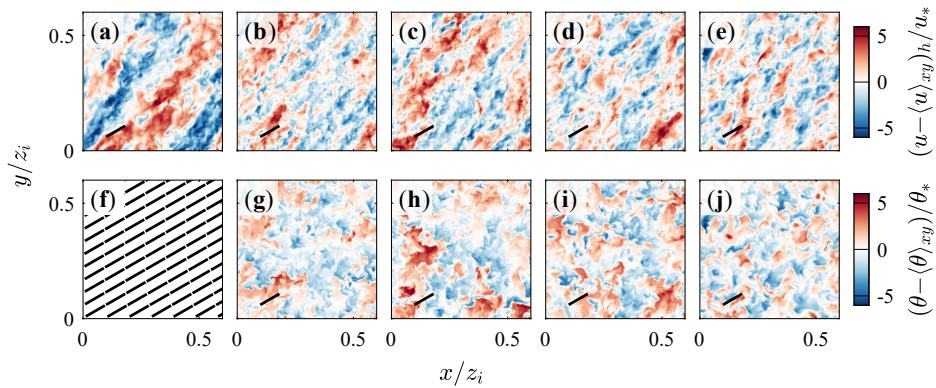}
\caption{Fluctuations of the horizontal velocity magnitude (top row) and temperature (bottom row) in the horizontal $x{-}y$ plane at $z=$ 0.05$z_i$. \textbf{a},\textbf{f} $z_i/L = 0$; \textbf{b},\textbf{g} $z_i/L = 1.7$; \textbf{c},\textbf{h} $z_i/L = 2.4$; \textbf{d},\textbf{i} $z_i/L = 3.2$; \textbf{e},\textbf{j} $z_i/L = 6.0$. The black rectangle in each panel represents the size of a local area aligned with $x^\prime$ that is used for the zone detection. An array of local detection areas is provided for reference in \textbf{f} rather than the temperature fluctuations due to negligible $\theta_*$ in this case.}
\label{fig:xy_fields}     
\end{figure}

The extent of UMZs in the transverse direction ($y^\prime$) is not known \textit{a priori} for the analysis. The distance $\mathcal{L}_{y^\prime}=0.01z_i$ used here for the width of the local flow volume is expected to be small relative to the width of the coherent regions. The horizontal area represented by $\mathcal{L}_{x^\prime,y^\prime}$ is shown in Fig. \ref{fig:xy_fields} in the context of near-surface turbulent fluctuations for each stability case. The horizontal area used for zone detection is smaller than or comparable to the coherent regions of velocity and temperature in each case. Hence, the choice of $\mathcal{L}_{x^\prime,y^\prime}$ is expected to manifest distinct histogram peaks for these regions. Trends in the apparent coherence at scales larger than the detection area are discussed later.

The number of grid points within each local flow volume is determined predominately by $z_i$ due to the scaling of $\mathcal{L}_{x^\prime,y^\prime}$. To mitigate the influence of different resolutions on the statistical convergence of the histograms, the data volumes are resampled to a matched grid spacing $\Delta = 0.003z_i$ for each case. This value corresponds to the original resolution of the neutrally-stratified LES, a 52\% increase in $\Delta$ for the $z_i/L=$ 1.7 case, and an 18\% increase for $z_i/L=$ 6.0. The final consideration for the local flow volume is its depth, which is limited here to the center of the low-level jet where the wind speed is maximized.

\begin{figure}
\centering
  \includegraphics[width=\linewidth]{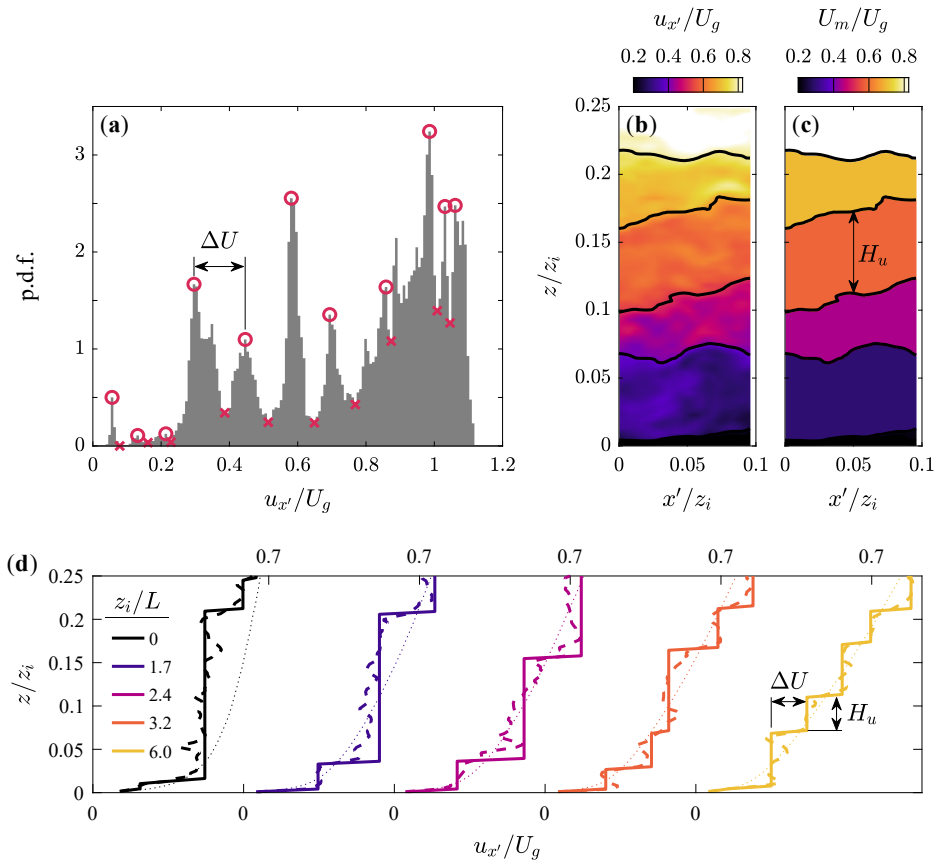}
\caption{Example detection of uniform momentum zones (UMZs). \textbf{a} Local histogram of the along-wind velocity component $u_{x^\prime}$, where each peak (circles) corresponds to the modal velocity $U_m$ of a quasi-uniform zone and the minima (crosses) correspond to the zone boundary. \textbf{b} Velocity field overlaid with isocontours of the velocity minima in (\textbf{a}). \textbf{c} The modal representation $U_m$ of the same velocity field. \textbf{d} Example instantaneous vertical profiles for each stability case comparing an LES velocity $u_{x^\prime}$ (dashed lines), its modal representation $U_m$ (solid lines), and the ensemble average $\langle u_{x^\prime} \rangle$ (dotted lines). The example in (\textbf{a,b,c}) is from the most stable case with $z_i/L = 6.0$.}
\label{fig:umz_detection}     
\end{figure}

Figure \ref{fig:umz_detection}a shows an example histogram computed using values of the resampled velocity $u_{x^\prime}$ within a volume confined by $\mathcal{L}_{x^\prime,y^\prime}$ and the jet height. The histogram is normalized as a probability density function (p.d.f.). An $x^\prime{-}z$ plane from the same local flow volume is presented in Fig. \ref{fig:umz_detection}b, where only the lowest 25\% of the PBL is shown to emphasize the visual signature of UMZs in the region of interest. The four large UMZs in Fig. \ref{fig:umz_detection}b correspond to four distinct peaks between 0.3$U_g$ and 0.7$U_g$ in Fig. \ref{fig:umz_detection}a. The histogram peaks are noticeably less definitive for higher velocities associated with positions farther from the surface. This behavior is attributed in part to Coriolis effects and directional shear. The rotated velocity component $u_{x^\prime}$ and narrow local flow volume become significantly misaligned with the mean wind direction far from the surface, which is another reason the analysis is limited to the lowest portion of the PBL.

In addition to the definition of the local flow volume, the width of the discrete histogram bins affects the apparent statistical convergence and peaks resulting from the histogram. The selected bin width 0.3$u_*$ is similar to values employed in previous studies \citep{Heisel2018,Heisel2020}. The bin width was determined by evaluating how the number of detected histogram peaks changed as the width decreased and choosing a width after the number of peaks became relatively invariant. Repeating the entire analysis for an alternate bin scaling 0.01$U_g$ \citep{deSilva2016} did not change the conclusions of the study.

The next step in the UMZ detection is defining a peak. Threshold parameters are used to determine whether a local maximum in the histogram is detected as a peak or assumed to be spurious. The parameters include the minimum peak prominence that measures the height of the peak above the neighboring local minima \citep{Laskari2018}. Here, a given local maximum must be 15\% greater than the adjacent local minima to be identified as a peak. In the Fig. \ref{fig:umz_detection}a example, several maxima at 0.35$U_g$ and above 0.8$U_g$ are excluded due to this threshold. The area of the peak is calculated as the integral of the p.d.f. between the local minima adjacent to the peak. This area corresponds to the number of grid points associated with the UMZ and thus the size of the UMZ. A minimum threshold of 0.01 (i.e. 1\% of both the p.d.f. and the local flow volume) is applied to the peak detection. The appropriate threshold parameters can vary between studies based on the data resolution and presence of measurement noise. The most important detail for the present study is the consistency of the local flow volume, resolution, and detection parameters across the cases considered.

The detected peaks resulting from the above parameters are indicated by circle markers in Fig. \ref{fig:umz_detection}a. Each peak indicates the presence of a UMZ and its representative modal velocity $U_m$ \citep{deSilva2016}. The thin shear layers along the UMZ edges span a relatively large range of $u_{x^\prime}$ across a small distance, resulting in a limited number of grid points corresponding to the velocity along the shear layer. The representative velocity of the UMZ edges can therefore be defined using the local minima between detected peaks \citep{Heisel2018}, shown as ``$\times$'' markers in Fig. \ref{fig:umz_detection}a. The spatial position of the UMZ edges, e.g. the black lines in Fig. \ref{fig:umz_detection}b, are estimated using isocontours of the representative edge velocity.

The turbulent field can be approximated as a series of UMZs by assuming perfect mixing $u_{x^\prime}=U_m$ throughout the UMZ extent as shown in Fig. \ref{fig:umz_detection}c. The instantaneous vertical profile of velocity corresponds to a single column of Fig. \ref{fig:umz_detection}b and \ref{fig:umz_detection}c. An example instantaneous profile for each flow case is provided in Fig. \ref{fig:umz_detection}d. The organization of the turbulent field into UMZs and thin shear layers yields a stairstep shape in the instantaneous profiles \citep{deSilva2016}, where the approximately constant rises correspond to UMZs and the steps occur across the thin shear layers. The instantaneous profile shape is distinct from the smooth mean (dotted lines), indicating the mean is only achieved through long-term averaging and variability in the position of the steps across space and time \citep{deSilva2017,Heisel2020}.

The approximation of the profiles as a series of instantaneous steps appears reasonably accurate for each stability case in Fig. \ref{fig:umz_detection}d. A quantitative assessment of the approximation is provided later in Sec. \ref{subsec:gradients}. The first relevant property of the UMZ organization and the stairstep profiles is the difference $\Delta U$ in modal velocity between adjacent UMZs defined as the difference between p.d.f. modes in Fig. \ref{fig:umz_detection}a. The velocity $\Delta U$ is also the approximate ``jump'' in $u_{x^\prime}$ across the thin shear layers in Fig. \ref{fig:umz_detection}d. The second property is the thickness $H_u$ of the UMZs calculated as the vertical distance between edges as seen in Figs. \ref{fig:umz_detection}c and \ref{fig:umz_detection}d. Both $\Delta U$ and $H_u$ are compiled for each $x^\prime$ position across the length $\mathcal{L}_{x^\prime}$ of the volume.

Due to the convoluted shape of the zone edges, the edges often cross a given $x^\prime$ position multiple times, e.g. near the coordinates ($x^\prime {=} 0.04z_i$, $z {=} 0.11z_i$) in Fig. \ref{fig:umz_detection}c, such that the zone appears as separated segments in the instantaneous profiles. In these cases, $H_u$ is calculated from the total thickness of the zone rather than from each individual segment. Additionally, a single $\Delta U$ is considered for the edge rather than including each crossing at the given $x^\prime$.

Figure \ref{fig:umz_detection} demonstrates the histogram-based detection of UMZs for a single local flow volume. The process is repeated for a new volume until the entire LES domain is analyzed for a given output instantaneous flow realization. The arrangement of local flow volumes is shown in Fig. \ref{fig:xy_fields}f, where the spacing between volumes along $y^\prime$ is 0.05$z_i$.

Zone statistics are not shown for the bottom 5\% of the PBL in later results. The grid resolution becomes coarse relative to the vertical extent of the turbulent features in this region, leading to issues in the zone detection and in the relevance of the zonal approximation at these heights. Many of the statistics deviate strongly below 0.05$z_i$ and are not considered representative of physical trends in the turbulence.

\subsection{Uniform temperature zone detection}
\label{subsec:detection_utz}

The detection of UTZs follows the same principles as Sec. \ref{subsec:detection} above. An account of similarities and differences between the UMZ and UTZ detection is given here.

The same local flow volume with horizontal area defined by $\mathcal{L}_{x^\prime,y^\prime}$, as visualized in Fig. \ref{fig:xy_fields}, is used to compute histograms of the turbulent temperature field. The resolution $\Delta = 0.003z_i$ is also fixed across flow cases. Rather than limiting the depth of the local flow volume based on the low-level jet, the region up to $z_i$ is included in calculations of the temperature histograms.

\begin{figure}
\centering
  \includegraphics[width=\linewidth]{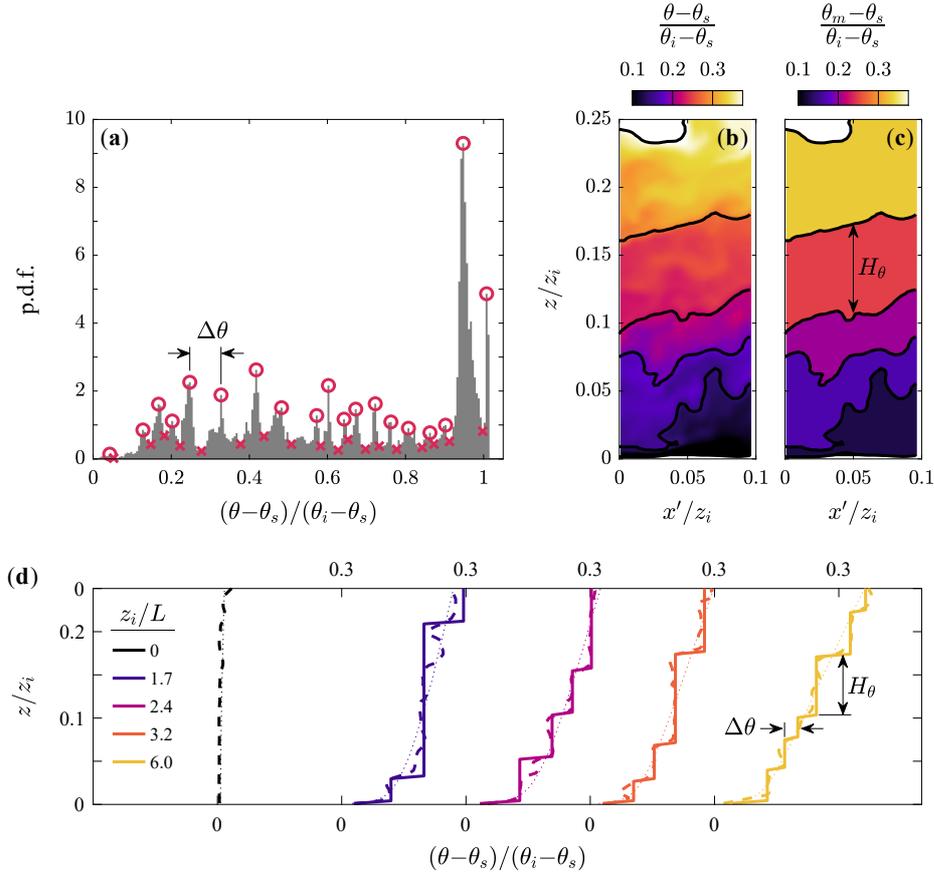}
\caption{Example detection of uniform temperature zones (UTZs) showing the same local flow field as Fig. \ref{fig:umz_detection}. \textbf{a} Local histogram of temperature, where each peak (circles) corresponds to the modal temperature $\theta_m$ of a quasi-uniform zone and the minima (crosses) correspond to the zone boundary. \textbf{b} Temperature field overlaid with isocontours of the temperature minima in (\textbf{a}). \textbf{c} The modal representation $\theta_m$ of the same temperature field. \textbf{d} Example instantaneous vertical profiles for each stability case comparing an LES temperature $\theta$ (dashed lines), its modal representation $\theta_m$ (solid lines), and the ensemble average $\langle \theta \rangle$ (dotted lines).}
\label{fig:utz_detection}     
\end{figure}

Figure \ref{fig:utz_detection}a shows a histogram of $\theta$ for the same local flow volume as the UMZ example in Fig. \ref{fig:umz_detection}. The bin width is 0.3$\theta_*$, matching the surface scaling used for the velocity histogram bins. The histogram peaks and UTZs are detected using the same thresholds as before for the relative prominence (15\%) and minimum area (0.01). The magnitude of the near-surface p.d.f. peaks is generally between 1 and 3 for both the velocity and temperature example histograms. As seen in Fig. \ref{fig:fields}, the temperature field is significantly more layered in the vicinity of the low-level jet, leading to a greater quantity of smaller histogram peaks away from the surface value.

As before, the representative temperature of each UTZ is the modal (peak) value $\theta_m$, and the edges are detected using isocontours of the minima between histogram peaks. The UTZs and their edges in Figs. \ref{fig:utz_detection}b and \ref{fig:utz_detection}c overlap significantly with the UMZs in Figs. \ref{fig:umz_detection}b and \ref{fig:umz_detection}c, specifically the edges located near $z=0.1z_i$ and 0.17$z_i$. There are also differences, notably for the near-surface zones. The example is emblematic of the turbulent velocity and temperature structure being closely, but not perfectly related.

The approximation of UTZs throughout the bottom 25\% of the PBL leads to the familiar stairstep-shaped instantaneous profiles in Fig. \ref{fig:utz_detection}d, where the steps are associated with temperature fronts \citep{Sullivan2016}. No large-scale organization of turbulent temperature eddies is expected for the neutrally-stratified flow, and UTZs are not detected for this case. The UTZ properties for the remaining cases are characterized in terms of the temperature difference $\Delta \theta$ between adjacent zones and the vertical thickness $H_\theta$ of the zones. Statistics for $\Delta \theta$ and $H_\theta$ are compiled for instances throughout the flow volume in the same manner as for the UMZs.

\subsection{Methodology summary}
\label{subsec:summary}

The LES of the PBL \citep{Sullivan2016} is based on the GABLS benchmark study \citep{Beare2006}. The LES includes five stability conditions ranging from neutral to near z-less stratification to explore the parameter space of interest. The key scaling parameters for each case are listed in Table \ref{tab1}.

Figure \ref{fig:mean_profiles} shows average profiles of the horizontal wind speed and temperature. Deviations in the profiles from the log law scaling parameters under increasing stratification are evident from the gradients in Figs. \ref{fig:mean_profiles}c and \ref{fig:mean_profiles}d. The focus of the study is how these deviations are related to properties of the instantaneous turbulence organization, and how these properties are accounted for in M-O similarity relations such as in Figs. \ref{fig:mean_profiles}e and \ref{fig:mean_profiles}f.

The near-surface coherent regions in the instantaneous turbulent flow, including elongated momentum streaks, are aligned with the surface wind direction that differs from the geostrophic wind direction due to Coriolis effects. These near-surface features are visualized in Fig. \ref{fig:streaks}. To align the analysis with these features, the horizontal plane of the LES output flow volumes is rotated according to the surface wind direction. The rotated coordinate system is given the notation ($x^\prime$,$y^\prime$) as seen in Fig. \ref{fig:streaks}.

The coherent regions in Fig. \ref{fig:streaks} are also apparent in the $x^\prime{-}z$ plane as seen in Fig. \ref{fig:fields}, where the layering of the regions creates the appearance of uniform zones separated by thin edges with enhanced gradients. Each uniform zone creates a distinct peak in histograms of the local flow volume. The representative value of the zone is given by the peak mode, and the zone edges are indicated by the minima between peaks. Detected modes and minima for an example histogram are shown in Fig. \ref{fig:umz_detection}a for UMZs and Fig. \ref{fig:utz_detection}a for UTZs. The corresponding zones in the $x^\prime{-}z$ plane are also shown in Figs. \ref{fig:umz_detection} and \ref{fig:utz_detection} for UMZs and UTZs, respectively, where the edge positions are estimated from isocontours of the detected minima.

Figures \ref{fig:umz_detection}d and \ref{fig:utz_detection}d show instantaneous profiles of the velocity and temperature, respectively, where the organization of uniform zones and thin gradient layers creates a stairstep-like shape in the profiles. The rises and steps are well approximated by the detected uniform zones for each flow case, such that the UMZ and UTZ properties can be used to relate the instantaneous turbulent structure to the smooth mean profile that results from variability in the position of the ``steps'' across long averaging periods. The zone properties are characterized here as the velocity difference $\Delta U$ between adjacent UMZs, the vertical thickness $H_u$ of UMZs, the temperature difference $\Delta \theta$ between adjacent UTZs, and the vertical thickness $H_\theta$ of UTZs.

Section \ref{sec:results} presents average statistics for these zone properties at heights between 0.05$z_i$ and 0.25$z_i$. At higher positions, the analysis is influenced by the behavior of the low-level jet and directional shear from Coriolis forces. Results are not shown for the bottom 5\% of the PBL because limitations in the LES and uniform zone methodology bias the statistics at the lowest positions.

\section{Results}
\label{sec:results}

\subsection{Applicability of the uniform zone approximation}
\label{subsec:gradients}

The two foremost properties for the organization of UMZs and UTZs are the uniformity of the flow within each zone and the clustering of gradients along the zone edges. Both properties are quantitatively assessed here using the detected zones and edges. The goal is to determine whether the zonal organization of turbulent eddies is present for the stably-stratified surface layer before proceeding to the evaluation of zone properties.

\begin{figure}
\centering
  \includegraphics[width=\linewidth]{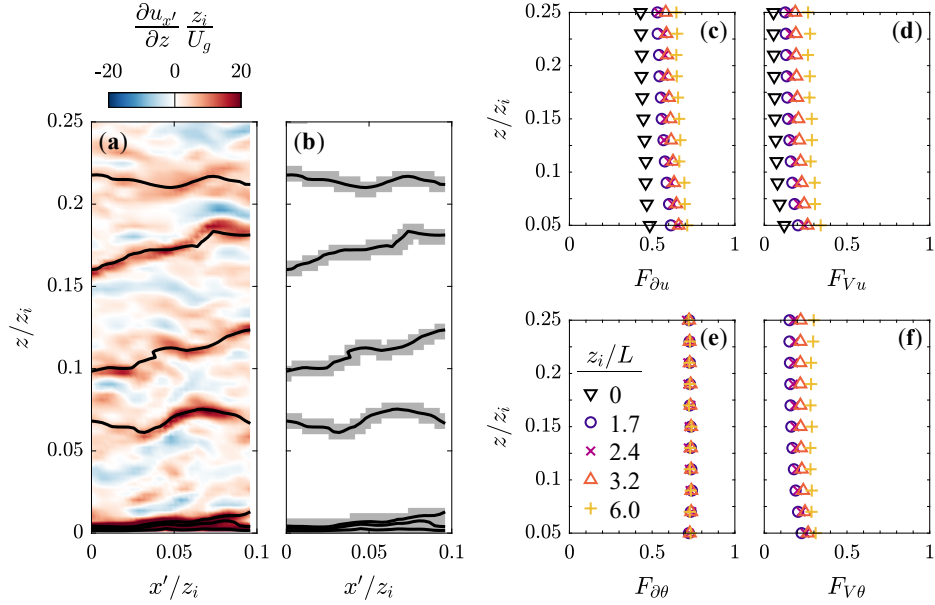}
\caption{Concentration of gradients along detected zone edges. \textbf{a} Velocity gradient $\partial u_{x^\prime} / \partial z$ for the example field in Fig. \ref{fig:umz_detection}b. \textbf{b} LES grid points along the zone edge, assuming the edge thickness is 0.01$z_i$. \textbf{c} Fraction of the velocity gradient $F_{\partial u} = (\Sigma \partial u_{x^\prime} / \partial z \vert_{edge}) / (\Sigma \partial u_{x^\prime} / \partial z)$ corresponding to the UMZ edges, computed in binned intervals of height $z$. \textbf{d} Fraction of the flow volume $F_{Vu}$ corresponding to the UMZ edges. \textbf{e} Fraction of the temperature gradient $F_{\partial \theta} = (\Sigma \partial \theta / \partial z \vert_{edge}) / (\Sigma \partial \theta / \partial z)$ corresponding to the UTZ edges. \textbf{f} Fraction of the flow volume $F_{V\theta}$ corresponding to the UTZ edges.}
\label{fig:fractions}     
\end{figure}

If small-scale statistics including vorticity, dissipation, and shear $\partial u_{x^\prime} / \partial z$ are spatially intermittent and preferentially aligned with the UMZ edges, statistics for $\partial u_{x^\prime} / \partial z$ computed at points along the edges will contribute disproportionately to the overall mean shear. The alignment can therefore be considered preferential if the contribution of the UMZ edges to $\partial u_{x^\prime} / \partial z$ is large relative to the fraction of the flow volume represented by the edges \citep{deSilva2017}. Figure \ref{fig:fractions}a shows an example instantaneous flow field demonstrating the overlap between the high-shear regions and the detected UMZ edges.

The edge thickness must be known or assumed to determine which grid points are associated with each UMZ edge. As discussed in Sec. \ref{subsec:orientation}, the effective thickness is expected to depend on the resolution $\Delta$ and the eddy viscosity of the LES. A thickness 0.01$z_i \approx 3\Delta$ is assumed here such that each edge spans three points on the interpolated grid. Note that moderately increasing or decreasing the assumed thickness does not change the conclusions drawn from the exercise. The points associated with the UMZ edges in Fig. \ref{fig:fractions}a are identified in \ref{fig:fractions}b for this assumed thickness.

The fractional contribution of these edge points to the shear statistics is $F_{\partial u} = (\Sigma \partial u_{x^\prime} / \partial z \vert_{edge}) / (\Sigma \partial u_{x^\prime} / \partial z)$, where the numerator is the shear only for points along the edges and the denominator is equivalent to the mean gradient. The fraction is computed height-by-height to yield the profile shown in Fig. \ref{fig:fractions}c. In each case, the instantaneous shear aligned with the UMZ edges accounts for 50-70\% of the overall mean. The fraction of the volume $F_{Vu}$ is estimated simply as the number of points along the edges divided by the total number of grid points at the given height. The volume fraction in Fig. \ref{fig:fractions}d is less than 30\% for each case. Results for the two fractions $F_{\partial u}$ and $F_{Vu}$ confirm the clustering of instantaneous shear along the UMZ edges: the detected edges are a majority contributor to the mean shear despite occupying a small fraction of the flow volume. This trend continues beyond the plotted limits shown until $z/h\approx$ 2/3, whereupon $F_{\partial u}$ and $F_{Vu}$ both decrease significantly.

The same edge statistics are presented for the UTZs, namely the fractional contribution $F_{\partial \theta}$ to the temperature gradient in Fig. \ref{fig:fractions}e and the fraction of the volume $F_{V\theta}$ in \ref{fig:fractions}f. The volume fraction is approximately the same as for the UMZs, but the contribution to the mean temperature gradient exceeds 70\%. The higher values for $F_{\partial \theta}$ are consistent with visual observations in Fig. \ref{fig:fields} that the UTZ edges (i.e. temperature fronts) are more distinct than the UMZ edges.

\begin{figure}
\centering
  \includegraphics[width=\linewidth]{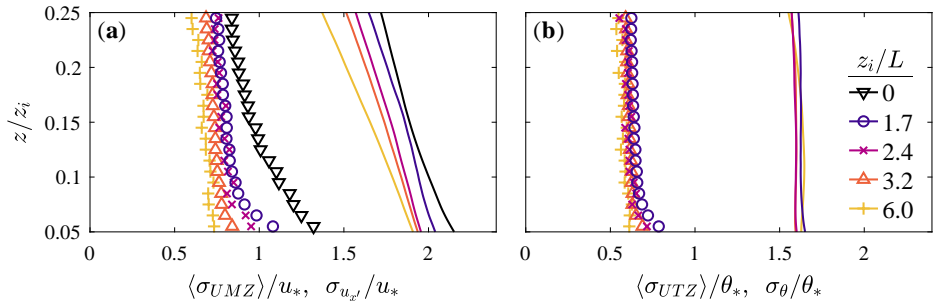}
\caption{Demonstration of zone uniformity using the root mean square (r.m.s.) of velocity and temperature within the detected zones. \textbf{a} The r.m.s. $\sigma_{UMZ}$ of velocity within a UMZ, averaged across zones at a given height. \textbf{b} The r.m.s. $\sigma_{UTZ}$ of temperature within a UTZ, averaged across zones at a given height. The zone r.m.s. statistics (symbols) are compared to the ensemble r.m.s. profiles $\sigma_{u_{x^\prime},\theta}$ (lines).}
\label{fig:rms}     
\end{figure}

Previous studies have demonstrated the uniformity of detected UMZs by showing the variability within each zone to be relatively small \citep{deSilva2016,Heisel2018}. The variability is measured here as the root mean square (r.m.s.) of velocity values across all points within a detected UMZ, excluding the edge points visualized in Fig. \ref{fig:fractions}b. The r.m.s. values are then averaged for all UMZs at a given position based on the UMZ midheight, resulting in the profiles $\langle \sigma_{UMZ} \rangle (z)$ shown in Fig. \ref{fig:rms}a. Variability within the UMZs is less than half of the overall r.m.s. of the LES velocity $\sigma_{u_{x^\prime}}$ that is also included in Fig. \ref{fig:rms}a. The same variability statistics for UTZs are shown in Fig. \ref{fig:rms}b, where $\langle \sigma_{UMZ} \rangle$ is almost three times smaller than $\sigma_{\theta}$.

The zone variability and LES r.m.s. are not directly comparable, because unlike for $\sigma_{u_{x^\prime},\theta}$, the r.m.s. within a single zone is computed across a range of heights spanned by that zone. Calculating the r.m.s. height-by-height within each zone yields an even greater reduction in the variability compared to $\sigma_{u_{x^\prime},\theta}$. The variance resulting statistically from the turbulent fluctuations is therefore predominately due to the passage of numerous uniform zones rather than the fluctuations within each zone. The residual variability within the UMZs and UTZs is attributed to relatively weak space-filling fluctuations whose organization (or lack thereof) does not contribute directly to the mean flow statistics but is closely related to scale-dependent trends \citep{Heisel2022}.

The results suggest that turbulence in the lowest portion of the stably-stratified PBL is approximately organized as a series of uniform zones (Fig. \ref{fig:rms}) and thin layers aligned with the largest gradients (Fig. \ref{fig:fractions}). The statistics support the visual evidence in Figs. \ref{fig:umz_detection}d and \ref{fig:utz_detection}d. Further, there is no indication that the organization weakens with increasing stratification, at least within the fully turbulent regime in the absence of global intermittency. The stability trend for the volume fraction in Figs. \ref{fig:fractions}d and \ref{fig:fractions}f is related to the zone thicknesses presented later. The smaller gradient contribution in Fig. \ref{fig:fractions}c and larger variability in Fig. \ref{fig:rms}a for the neutral case may be related to the coarser native resolution $\Delta/z_i$ of the simulation and its effective eddy viscosity as previously discussed.

\subsection{Zone velocity and temperature}

\begin{figure}
\centering
  \includegraphics[width=\linewidth]{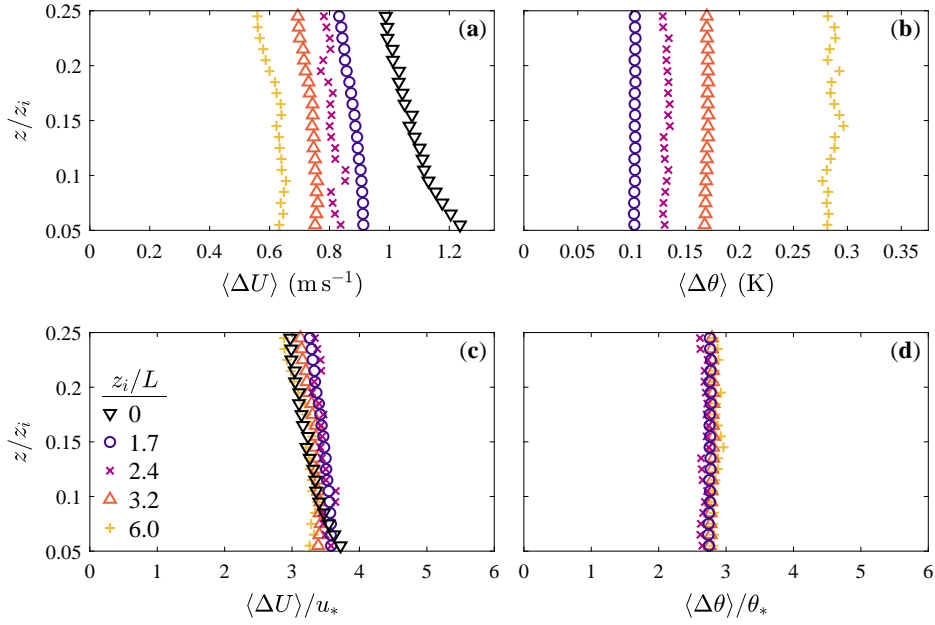}
\caption{Profiles of the mean difference in velocity and temperature between vertically adjacent zones. \textbf{a,c} Velocity difference between UMZs $\langle \Delta U \rangle$ as defined in Fig. \ref{fig:umz_detection}. \textbf{b,d} Temperature difference between UTZs $\langle \Delta \theta \rangle$ as defined in Fig. \ref{fig:utz_detection}. The profiles are shown dimensionally and relative to the surface scaling parameters.}
\label{fig:jump_profiles}     
\end{figure}

The characteristic velocity $\Delta U$ and temperature $\Delta \theta$ of detected zones correspond to sharp changes or ``jumps'' in value across the gradient layers as seen in Figs. \ref{fig:umz_detection}d and \ref{fig:utz_detection}d. The compiled statistics for $\Delta U$ and $\Delta \theta$ are averaged here in binned intervals of height based on the position of the zone edge where the jump occurs. Profiles of the binned averages are shown in Fig. \ref{fig:jump_profiles}.

The profiles are shown dimensionally in Figs. \ref{fig:jump_profiles}a and \ref{fig:jump_profiles}b to demonstrate the clear stability trends, and are plotted relative to the surface parameters $u_*$ and $\theta_*$ in Figs. \ref{fig:jump_profiles}c and \ref{fig:jump_profiles}d, respectively. The surface parameters yield agreement across all cases for both $\Delta U$ and $\Delta \theta$.

The result $\Delta U \sim u_*$ and the moderate decrease in $\Delta U(z)$ with height are both consistent with previous studies \citep{deSilva2017,Heisel2020}, including field measurements in the neutrally-stratified surface layer \citep{Heisel2018}. In canonical turbulent boundary layers, the relation $\Delta U \sim u_*$ extends to the top of the boundary layer \citep{Heisel2020}. It is not known if $\Delta U$ is invariant with height in a true logarithmic region near to the surface. As discussed, the resolution becomes a limiting factor in the lowest portion of the boundary layer, both for the present LES and for the previous experimental measurements referenced above. The UMZ detection and $\Delta U$ calculations were repeated using the magnitude of the horizontal velocity rather than the rotated component $u_{x^\prime}$. The velocity $\Delta U$ was modestly larger in this case, but the same height dependence and agreement across cases for $\Delta U / u_*$ were observed.  The directional shear therefore does not change the primary trends in $\Delta U$ below 0.25$z_i$.

In contrast to the velocity, the temperature $\Delta \theta$ is approximately constant with height in Fig. \ref{fig:jump_profiles}d. The magnitude of the difference relative to the surface parameters is moderately smaller for $\Delta \theta$ than $\Delta U$. Considering the same bin width (i.e. 0.3$u_*$ and 0.3$\theta_*$) was employed in the zone detection, the differing intensity may be a physical characteristic of the turbulence. The trend is further discussed in Sec. \ref{subsec:prandtl}.

The results in Figs. \ref{fig:jump_profiles}c and \ref{fig:jump_profiles}d indicate that the characteristic intensities $\Delta U$ and $\Delta \theta$ of the persistent eddy organization maintain a close proportional dependence on the surface fluxes regardless of the stratification. The surface properties $u_*$ and $\theta_*$ are the relevant scaling  parameters for the logarithmic profiles $\partial U / \partial z = u_*/\kappa z$ and $\partial \theta / \partial z = \theta_*/\kappa z$ such that the eddy intensity is consistent with the log law even under stratified conditions. The result is unsurprising considering the absence of alternate velocity scales such as $w_*$ that becomes relevant for convective conditions. For stable stratification, the turbulence is shear-driven such that the generation of variability (i.e. turbulent energy) in the velocity originates solely from a surface shear stress boundary condition and $u_*$. Likewise, variability in the temperature is closely linked to the surface heat flux boundary condition and $\theta_*$.

\subsection{Zone thickness}

\begin{figure}
\centering
  \includegraphics[width=\linewidth]{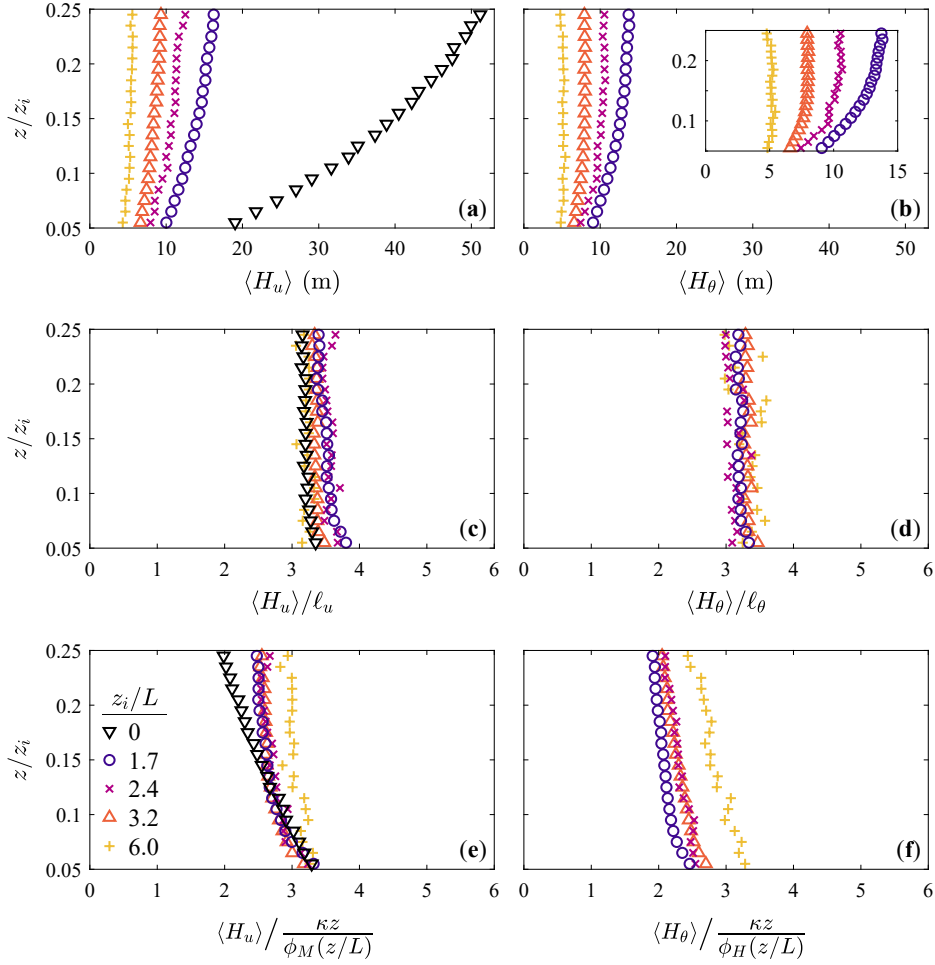}
\caption{Profiles of the mean zone thickness. UMZ thickness $\langle H_u \rangle$ as defined in Fig. \ref{fig:umz_detection}, shown dimensionally (\textbf{a}), relative to the length $\ell_u$ in Eq. \ref{eq:Lulocal} (\textbf{c}), and relative to the similarity relation for momentum $\phi_M$ (\textbf{e}). UTZ thickness $\langle H_\theta \rangle$ as defined in Fig. \ref{fig:utz_detection}, shown dimensionally (\textbf{b}), relative to the length $\ell_\theta$ in Eq. \ref{eq:Lt} (\textbf{d}), and relative to the similarity relation for heat $\phi_H$ (\textbf{f}).}
\label{fig:thickness_profiles}     
\end{figure}

Given the agreement of $\Delta U$ and $\Delta \theta$ with the respective log law parameters $u_*$ and $\theta_*$ in Fig. \ref{fig:jump_profiles}, deviations from the logarithmic profiles must predominately correspond to changes in the geometry of the uniform zones. The geometry is characterized in terms of the vertical zone thickness $H_{u,\theta}$ as shown in Figs. \ref{fig:umz_detection}d and \ref{fig:utz_detection}d. The compiled thickness statistics are averaged here in binned intervals of $z$, where the representative position of each zone is taken to be its midheight. Profiles of the average zone thickness are shown in Fig. \ref{fig:thickness_profiles}.

The dimensional profiles in Figs. \ref{fig:thickness_profiles}a and \ref{fig:thickness_profiles}b reveal the expected trends in both height and stability. For neutral stratification, the UMZ thickness appears to increase proportionally with height $z$, matching previous experimental observations \citep{Heisel2020}. The presence of weak stability leads to a sharp reduction in the zone thickness, consistent with the pancake-like structure observed for stratified turbulence \citep[e.g.][]{Caulfield2021}. In the $z_i/L=$ 1.7 case, a height dependence is still observed for both $H_u$ and $H_\theta$, but the zones are several times thinner than for neutral conditions.

The reduction in $H_{u,\theta}$ continues with increasing stratification. The smaller zones lead to more numerous ``steps'' and a stronger mean gradient in Figs. \ref{fig:umz_detection}d and \ref{fig:utz_detection}d. The increased number of steps also leads to the zone edges representing a larger fraction of the flow as seen in Figs. \ref{fig:fractions}d and \ref{fig:fractions}f. For $z_i/L=$ 6 in Figs. \ref{fig:thickness_profiles}a and \ref{fig:thickness_profiles}b, the thickness appears invariant with height relative to the neutral case, particularly for heights above 0.1$z_i$. The observed eddy geometry therefore spans the range from ``attached'' in accordance with log law predictions for the neutral PBL to relatively independent of height in accordance with the regime of z-less stratification.

For each LES case, an approximate local equilibrium between shear production, buoyancy damping, and dissipation is observed within the lowest 25\% of the PBL that is the target region of the study. The equilibrium can be used to account for the UMZ thickness trends in Figs. \ref{fig:thickness_profiles}a. Under these equilibrium conditions, i.e. for steady state flow with negligible transport by turbulence and pressure, the TKE budget is

\begin{equation}
0 = - \langle u^\prime w^\prime \rangle \frac{\partial \langle U \rangle }{\partial z} + \frac{g}{\langle \theta \rangle} \langle w^\prime \theta^\prime \rangle - \epsilon,
\label{eq:tkebudget}
\end{equation}

\noindent where the right-hand side terms respectively represent shear production of turbulence, production or suppression of turbulence by buoyancy, and the rate of dissipation $\epsilon$ by viscosity. The observed spatial organization of eddies and the corresponding stairstep-like instantaneous profiles support the deconstruction of the average gradients using the local zone properties \citep{Heisel2020}:

\begin{equation}
\frac{\partial  \langle U \rangle }{\partial z} \approx \frac{\langle \Delta U \rangle}{\langle H_u \rangle} \sim \frac{ u_* }{\ell_u} \quad \mathrm{and} \quad \frac{\partial  \langle \theta \rangle }{\partial z} \approx \frac{\langle \Delta \theta \rangle}{\langle H_\theta \rangle} \sim \frac{ \theta_* }{\ell_\theta}.
\label{eq:gradients}
\end{equation}

\noindent Here, $\Delta U \sim u_*$ and $\Delta \theta \sim \theta_*$ are supported by Fig. \ref{fig:jump_profiles}, and the lengths $\ell_u$ and $\ell_\theta$ are the parameters to be determined. Substituting for $\partial \langle U \rangle / \partial z \sim u_*/\ell_u$ in Eq. \ref{eq:tkebudget} and solving for $\ell_u$ yields the length scale

\begin{equation}
\ell_u = u_* \left( \frac{g \langle w^\prime \theta^\prime \rangle }{\langle \theta \rangle \langle u^\prime w^\prime \rangle }  - \frac{\epsilon}{\langle u^\prime w^\prime \rangle } \right)^{-1}.
\label{eq:Lulocal}
\end{equation}

\noindent Note that Eq. \ref{eq:Lulocal} is specific to neutral and stable conditions because the zonal organization and the relations in Eq. \ref{eq:gradients} have not been evaluated for the convective PBL.

Figure \ref{fig:thickness_profiles}c shows $H_u$ relative to $\ell_u$, where the dissipation term in $\ell_u$ is estimated using the model $\epsilon \approx C_\epsilon e^{3/2}/\Delta$ based on the SGS TKE $e$ and constant $C_\epsilon =$ 0.93 \citep{Moeng1988,Sullivan2016}. The agreement across cases and invariance with height validate the equilibrium assumed in Eq. \ref{eq:tkebudget} and confirm the connection between the gradient and zone properties in Eq. \ref{eq:gradients}. Further, the use of local fluxes in the definition of $\ell_u$ is consistent with the concept of local scaling for stable conditions \citep{Nieuwstadt1984}.

For the simplified budget in Eq. \ref{eq:tkebudget}, the relative contributions of the dissipation $\epsilon$ and buoyancy $B=\tfrac{g}{\langle \theta \rangle} \langle w^\prime \theta^\prime \rangle$ to $\ell_u$ are related directly to the flux Richardson number $R_f$ as $\epsilon/B = R_f^{-1}-1$. Considering the flux Richardson number has a critical maximum value $R_f \approx 0.2$ under stable stratification \citep{Yamada1975,Grachev2013,Bouzeid2018}, the buoyancy term is always at least four times smaller than the dissipation. The relatively small buoyancy contribution suggests that the approach to z-less stratification is predominately due to a reduced dependence on $z$ in both $\partial U / \partial z$ and $\epsilon$ rather than the direct contribution of buoyancy in Eq. \ref{eq:tkebudget}. The primary effect of the buoyancy term in Eq. \ref{eq:Lulocal} is a bulk reduction in $\ell_u$ that is proportional to $\theta_*/u_*$.

The same equilibrium assumption can be used to infer a length scale $\ell_\theta$ from the budget equation for temperature variance. The simplified budget for half the temperature variance is \citep{Stull1988,Mironov2016}

\begin{equation}
0 = - \frac{1}{2}\langle w^\prime \theta^\prime \rangle \frac{\partial \langle \theta \rangle }{\partial z} - \epsilon_\theta,
\label{eq:tempbudget}
\end{equation}

\noindent where $\epsilon_\theta$ is the molecular dissipation rate for temperature. Substituting for $\partial \langle \theta \rangle / \partial z \sim \theta_*/\ell_\theta$ from Eq. \ref{eq:gradients} into Eq. \ref{eq:tempbudget} and solving for $\ell_\theta$ yields

\begin{equation}
\ell_\theta = - \frac{ \theta_* \langle w^\prime \theta^\prime \rangle }{2 \epsilon_\theta}.
\label{eq:Lt}
\end{equation}

\noindent The dissipation $\epsilon_\theta$ is estimated here as $\epsilon_\theta \approx C_\theta e^{1/2} \theta_{sgs}^2 /\Delta$ using the constant $C_\theta \approx$ 2.1 \citep{Moeng1988}. For this analysis, the variance of the SGS temperature $\theta_{sgs}^2$ is approximated from the integral of the turbulent temperature spectrum extrapolated to infinite wavenumber.

Figure \ref{fig:thickness_profiles}d shows $H_\theta$ relative to $\ell_\theta$. The ratio $H_\theta/\ell_\theta$ exhibits the same magnitude and agreement across cases seen for $H_u/\ell_u$. The temperature variance budget contains no coupled term akin to the buoyancy damping in the TKE, such that the reduction in height dependence of $H_\theta$ and $\ell_\theta$ with increasing stratification corresponds solely to trends in $\epsilon_\theta (z)$.

While $\ell_{u,\theta}$ more accurately captures the zone thickness trends when defined using local fluxes, it is informative to apply the surface scaling assumptions $\langle u^\prime w^\prime \rangle = -u_*^2$ and $\langle w^\prime \theta^\prime \rangle = -u_* \theta_*$. Substituting for surface parameters in Eq. \ref{eq:Lulocal} and multiplying the numerator and denominator by $\kappa z$ yields

\begin{equation}
\ell_u \approx \kappa z \left( \frac{z }{L }  + \frac{\kappa z}{L_\epsilon } \right)^{-1}.
\label{eq:LuMOST}
\end{equation}

\noindent The length $L_\epsilon = u_*^3/\epsilon$ corresponds to the production range of scales in boundary layer flows \citep{Davidson2014,Chamecki2017,Ghannam2018}, and simplifies to $L_\epsilon \approx \kappa z$ in neutrally-stratified conditions when dissipation and shear production are both approximately $u_*^3/\kappa z$.

The definition of $\ell_u$ in Eq. \ref{eq:LuMOST} is similar in form to existing mixing length models \citep[e.g.][]{Blackadar1962,Delage1974,Huang2013a}. Additionally, the denominator can be interpreted to represent departure from the log law scaling predicted by $\phi_M$. As discussed previously, the loss of height dependence in $\ell_u$ predominately relates to $\kappa z / L_\epsilon$ diverging from unity, and is only modestly affected by the direct contribution of $z/L$. However, in traditional M-O relations the dissipative term $\kappa z / L_\epsilon$ is also parameterized in terms of $z/L$ as $\kappa z / L_\epsilon = \epsilon \kappa z / u_*^3 = \phi_\epsilon(z/L)$. The relation $\phi_M = \phi_\epsilon + z/L$ \citep{Hartogensis2005} results naturally from Eq. \ref{eq:LuMOST} and is consistent with $\kappa z / L_\epsilon$ representing a majority of the similarity correction that is captured by $\phi_M$.

The connection between $\phi_{M}$ and the UMZ geometry (through $\ell_u$) is evaluated in Fig. \ref{fig:thickness_profiles}e. The plot should match $H_u/\ell_u$ in Fig. \ref{fig:thickness_profiles}c if the surface scaling assumption is accurate and if $(z/L + \kappa z / L_\epsilon) \approx \phi_M = (1+4.7z/L)$. The same comparison for the UTZ thickness is shown in Fig. \ref{fig:thickness_profiles}f. In both cases, the M-O relations account for a majority of the trends in height and stability of zone thickness that are observed in Figs. \ref{fig:thickness_profiles}a and \ref{fig:thickness_profiles}b. However, there is a notable difference in the $z_i / L =$ 6 case and a moderate height-dependence that is likely related to the surface scaling assumption. The trends are consistent with the mean gradients in Figs. \ref{fig:mean_profiles}e and \ref{fig:mean_profiles}f, where $\phi_{M,H}$ accurately describes the bulk of the deviation from the log law in each stability case, but also exhibit residual differences with a possible stability trend.

The purpose of the lengths $\ell_{u,\theta}$ is to explore the inter-related nature of the eddy organization, the turbulence budget contributions, and M-O similarity relations. The contribution of turbulent transport is neglected here, but is relevant to measurements in the roughness layer and around complex terrain \citep{Heisel2020a,Chamecki2020}. The practical utility of the lengths is also limited by challenges in estimating $\epsilon$ and $\epsilon_\theta$. Further, the inclusion of dissipation in the definitions for $\ell_u$ and $\ell_\theta$ should not be interpreted as causality, e.g. $\ell_u = f(\epsilon,...)$. Rather, it is expected that the properties of the integral-scale turbulence (e.g. $\ell_{u,\theta}$) determine the required dissipation rate of the small scales in accordance with cascade theory \citep[e.g.][]{Pope2000}.

\subsection{Zone edge layers}

The previous statistics demonstrated the relevance of the surface parameters $u_*$ and $\theta_*$ and length scales $\ell_{u,\theta}$ to the bulk properties of the uniform zone organization of turbulent eddies. These properties are related to the stairsteps of the instantaneous profiles in Figs. \ref{fig:umz_detection}d and \ref{fig:utz_detection}d. The zone scaling parameters and the stairstep pattern are also evident from the average flow behavior around a zone edge, as demonstrated here.

\begin{figure}
\centering
  \includegraphics[width=\linewidth]{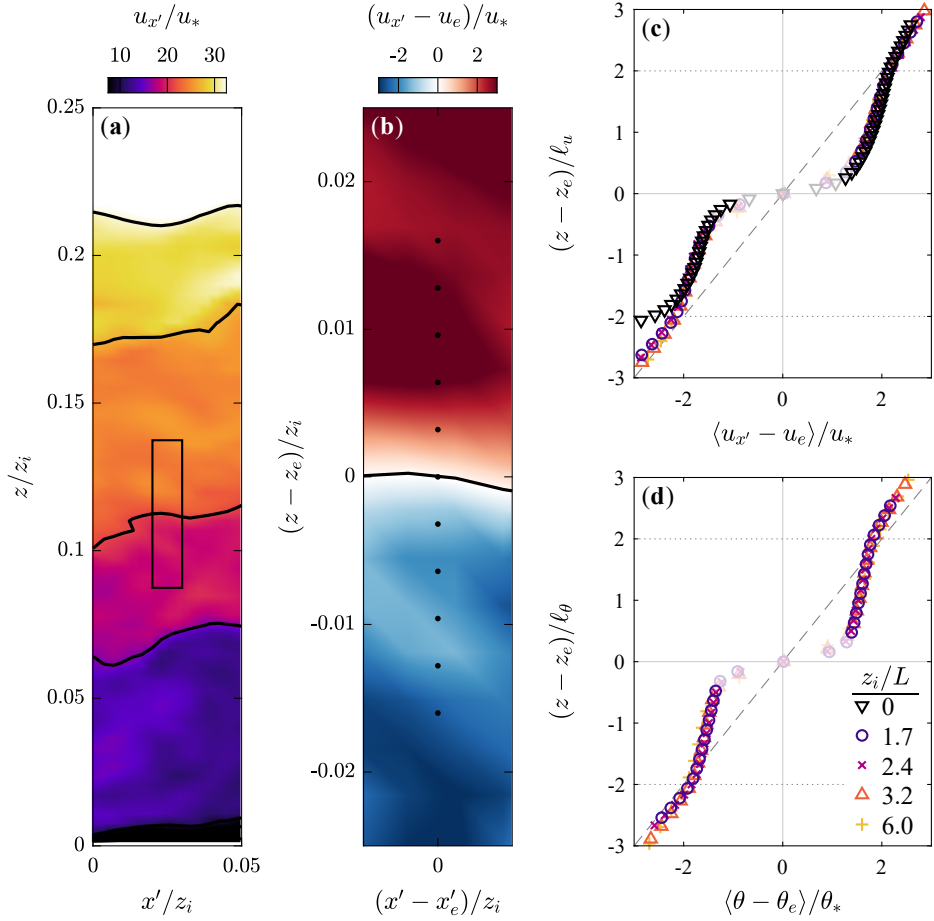}
\caption{Average profiles conditional to detected zone edges. \textbf{a} Example velocity field from Fig. \ref{fig:umz_detection}b. \textbf{b} Velocity in the vicinity of a detected UMZ edge, i.e. the boxed area in \textbf{a}, relative to the edge position $z_e$ and velocity $u_e$. \textbf{c} Velocity profiles relative to the detected edge. \textbf{d} Temperature profiles relative to the detected edge. The profiles are averaged across all edges at $z_e\approx 0.1z_i$, excluding instances where the nearest adjacent edge is within 2$\ell_{u , \theta}$, i.e. the dotted lines in \textbf{c} and \textbf{d}. The dashed line represents the average gradient at $z_e$. Local dynamics within the edge layers (shown as transparent markers) are expected to depend strongly on the subgrid-scale model of the simulations.}
\label{fig:interface_profiles}     
\end{figure}

Each detected zone edge has a position $(x_e,z_e)$ corresponding to an isocontour of the representative edge velocity $u_e$ or temperature $\theta_e$. The representative value for each isocontour is determined from the p.d.f. minima as discussed in Sects. \ref{subsec:detection} and \ref{subsec:detection_utz}. Figure \ref{fig:interface_profiles}a shows the zone edges (black lines) for an example field. The flow profile around a zone edge can be computed in a frame of reference relative to the edge properties, e.g. velocity $(u-u_e)$ as a function of distance from the edge center $(z-z_e)$ as shown in Fig. \ref{fig:interface_profiles}b. Profiles of velocity $(u-u_e)$ and temperature $(\theta-\theta_e)$ were compiled in this reference frame for every edge position, and the profiles were averaged across edges with position $z_e \approx 0.1z_i$. The same methodology for computing conditional edge profiles has been applied to experimental measurements \citep{deSilva2017,Heisel2021}.

Figs. \ref{fig:interface_profiles}c and \ref{fig:interface_profiles}d show the resulting averaged flow profiles centered around the detected zone edges. The collapse of the profiles provide further support for using $u_*$, $\theta_*$, and $\ell_{u,\theta}$ to characterize the layered organization of turbulence in the stratified PBL. The zone edges and statistics represented by Fig. \ref{fig:interface_profiles} were detected from a reanalysis with alternate histogram bin widths 0.01$U_g$ and 0.005($\theta_i-\theta_s$). The profiles suggest the scaling of the zone properties with $u_*$, $\theta_*$, and $\ell_{u,\theta}$ observed in previous figures is not an artifact of the original bin widths. A notable exception in the collapse is the lowest $z$ positions for the neutral case in Fig. \ref{fig:interface_profiles}c. The length $\ell_u$ varies strongly with $z$ in this case, such that the value $\ell_u(z_e)$ at the center of the reference frame is not applicable to statistics far from the center.

The transparent markers in Fig. \ref{fig:interface_profiles} represent the region in the immediate vicinity of $z_e$ where the dynamics depend strongly on the SGS model and grid resolution. For turbulent boundary layers, the thickness of the shear layers is proportional to the Taylor microscale \citep{Eisma2015,deSilva2017,Heisel2021}. The microscale depends on viscous processes that are not explicitly included in the present simulations. A majority of the jump in velocity and temperature occurs across this thin region, which is consistent with the preferential alignment of gradients in Fig. \ref{fig:fractions} and the jumps appearing as an approximate discontinuity in the instantaneous profiles.

The averages in Figs. \ref{fig:interface_profiles}c and \ref{fig:interface_profiles}d exclude instances in which the nearest adjacent edge is closer than 2$\ell_{u,\theta}$, such that no other zone edges occur within the region bounded by horizontal dotted lines. The heights within these limits and away from the center $z=z_e$ therefore represent the interior of the uniform zones. As expected, the velocity and temperature vary weakly within this region relative to the sharp change at $z=z_e$. The mean gradients from Eq. \ref{eq:gradients} are included as dashed lines in Fig. \ref{fig:interface_profiles}. The individual components of the organized structure, i.e. the uniform zones and their edges, both have shapes differing significantly from the mean gradient, but the combined properties of the average edge and adjacent zone(s) matches the mean gradient across the extent of the zone from $(z-z_e)=0$ to 2$\ell_{u,\theta}$.

Outside the dotted lines, the profiles approximately match the mean gradient. Neighboring zone edges can appear at any height within this region. The alignment with the mean gradient in this region is due to variability in the position of these neighboring edges across different instantaneous profiles that contribute to the average. This is in contrast to the stairstep shape within the dotted lines achieved by fixing the edge position in the average. The different behavior emphasizes the point that while the magnitude of the mean gradients corresponds closely to the existence and properties of the uniform zones, the continuity of the gradients with height (i.e. a smooth mean profile) depends on variability in the zone edge position across space and time.

\subsection{Uniform zones and the turbulent {P}randtl number}
\label{subsec:prandtl}

The analysis thus far has demonstrated the contribution of the size and intensity of the prevailing turbulent eddies to the mean gradients. The deconstruction of the gradient based on the size (i.e. $H_{u,\theta}$) and intensity (i.e. $\Delta U$ and $\Delta \theta$) is given in Eq. \ref{eq:gradients}. The same principle can be applied to the investigate the relation between the eddy organization and the turbulent Prandtl number:

\begin{equation}
Pr_t = \dfrac{ \langle u^\prime w^\prime \rangle_h \dfrac{\partial \langle \theta \rangle }{ \partial z} }{ \langle w^\prime \theta^\prime \rangle \dfrac{ \partial \langle U \rangle_h }{ \partial z } } \approx \dfrac{ \langle \Delta \theta \rangle / \langle w^\prime \theta^\prime \rangle }{ \langle \Delta U \rangle / \langle u^\prime w^\prime \rangle_h } \dfrac{ \langle H_u \rangle }{ \langle H_\theta \rangle }.
\label{eq:prandtl}
\end{equation}

\begin{figure}
\centering
  \includegraphics[width=\linewidth]{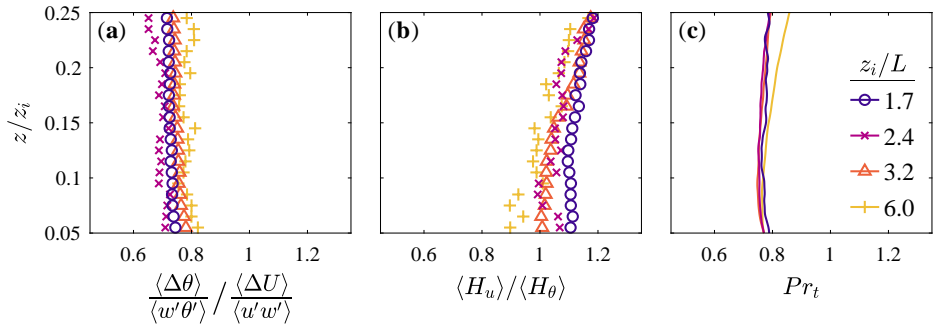}
\caption{Profiles illustrating the contribution of zone properties, i.e. eddy intensity and geometry, to trends in the turbulent Prandtl number. \textbf{a} Ratio of average differences $\Delta \theta$ and $\Delta U$ from Fig. \ref{fig:jump_profiles}, normalized by the local fluxes. \textbf{b} Ratio of average thicknesses $H_u$ and $H_\theta$ from Fig. \ref{fig:thickness_profiles}. \textbf{c} Turbulent Prandtl number $Pr_t = \left( -\langle u^\prime w^\prime \rangle_h \partial \langle \theta \rangle / \partial z \right) / \left( -\langle w^\prime \theta^\prime \rangle \partial \langle U \rangle_h / \partial z \right)$ corresponding to the product of \textbf{a} and \textbf{b}.}
\label{fig:prandtl_number}     
\end{figure}

\noindent The turbulent Prandtl number represents dissimilarity in the turbulent diffusivity of heat and momentum, and is a key parameter in modeling turbulence effects in the atmosphere \citep{Kays1994,Li2019}. Equation \ref{eq:prandtl} implies that $Pr_t$ depends on similarity in both the eddy intensity relative to the fluxes and in the eddy geometry. The components of Eq. \ref{eq:prandtl} are shown in Fig. \ref{fig:prandtl_number}.

The comparison of the average differences $\Delta U$ and $\Delta \theta$ in Fig. \ref{fig:prandtl_number}a reveals that $\Delta U$ is consistently larger than $\Delta \theta$ relative to the fluxes for momentum and heat. The same trend is apparent from the differing magnitudes in Figs. \ref{fig:jump_profiles}c and \ref{fig:jump_profiles}d. The height dependence of $\Delta U$ observed previously in Fig. \ref{fig:jump_profiles}c is significantly weaker in Fig. \ref{fig:prandtl_number}a. The weaker trend in Fig. \ref{fig:prandtl_number}a is due to $\Delta U$ having a similar height dependence as the ratio of the local flux profiles $\langle w^\prime \theta^\prime \rangle / \langle u^\prime w^\prime \rangle_h $, where the momentum flux decays faster in height than the approximately linear heat flux profile for stable conditions \citep{Nieuwstadt1984}. It is not clear from the present results whether there is a phenomenological connection between the height dependence of $\Delta U$ and the deviation from linear decay in the momentum flux profile.

The ratio of zone thicknesses $H_{u,\theta}$ in Fig. \ref{fig:prandtl_number}b indicates the UMZs and UTZs have similar thickness in the surface layer below 0.1$z_i$. The similar thicknesses are seen also in a comparison of Figs. \ref{fig:thickness_profiles}a and \ref{fig:thickness_profiles}b. The similarity is specific to the vertical extent of the zones, as the example fields in Fig. \ref{fig:xy_fields} indicate a level of dissimilarity in the horizontal extent. The thickness ratio deviates from unity away from the surface in Fig. \ref{fig:prandtl_number}b due to a weaker height dependence in the UTZ size that is also apparent in Fig. \ref{fig:thickness_profiles}b. The increase in $H_u/H_\theta$ becomes more pronounced at higher positions above the plotted limits. The trend is related to the previous observation from Fig. \ref{fig:fields} that the temperature field has more zones or ``layers'' compared to the velocity field in the upper portion of the PBL approaching the low-level jet.

As seen from Eq. \ref{eq:prandtl}, the Prandtl number in Fig. \ref{fig:prandtl_number}c results from the product of the ratios in panels a and b. The smaller diffusivity for heat, indicated by $Pr_t < 1$, can therefore be attributed to a weaker eddy intensity relative to the fluxes and surface parameters; the difference in temperature $\Delta \theta / \theta_*$ across the fronts is approximately 20\% smaller than $\Delta U/u_*$ across the shear layers. The Prandtl number is approximately constant near the surface where the vertical extent of the uniform zones is similar. However, $Pr_t$ increases with height above the limits of Fig. \ref{fig:prandtl_number} where the UTZs become thinner and more numerous compared to the UMZs.

The trends in Fig. \ref{fig:prandtl_number} also correspond to $\phi$ through the relation $Pr_t = \phi_M / \phi_H$. The M-O similarity relations used here follow a linear form $\phi = a+b(z/L)$. The difference in $a$ for momentum ($a=$1) and heat ($a=$ 0.74) represents a shift in the gradient profiles that correspond directly to the discrepancy between $\Delta \theta / \theta_*$ and $\Delta U/u_*$. The relation between zone properties in Fig. \ref{fig:prandtl_number}a and \ref{fig:prandtl_number}b does not vary with height below $z\approx 0.1z_i$, which corresponds to the similarity in $b$ for momentum and heat ($b=$ 4.7 for both) and the approximately constant $Pr_t$ within the surface layer.

\section{Concluding remarks}
\label{sec:conclusion}

As computational and experimental capabilities have advanced in recent decades, there is a growing body of evidence that small-scale intermittency (i.e. non-uniform distribution of dissipation in space and time) is readily apparent from instantaneous realizations of high-Reynolds-number turbulence \citep{Ishihara2009,Hunt2010}. This spatial intermittency results in visually striking patterns when the forcing conditions induce a preferential orientation in the clusters of small-scale eddies. For instance, the anisotropy for both boundary layer flows and stratified flows leads to distinct layers of elongated uniform regions separated by much thinner regions with intense gradients \citep{Meinhart1995,Caulfield2021}. In the boundary layer case, the thin gradient regions have a positive average inclination relative to the horizontal plane as seen in Fig. \ref{fig:fields}, leading to the signature of ramp-like structures in two-point correlation statistics \citep{Hutchins2012,Chauhan2013,Liu2017}.

It is perhaps unsurprising that the same layered features are observed here for the stable PBL which is both wall-bounded and stratified. The layers in each realization are detected as UMZs and UTZs, which assumes the flow within the zones is uniform and that all gradients coincide with the zone edges. The assumption is supported by conditional statistics showing the gradients are preferentially aligned with the detected zone edges (Fig. \ref{fig:fractions}) and the variability within the zones is small (Fig. \ref{fig:rms}). 

Many aspects of the complex turbulent field are lost when the flow is simplified to a series of UMZs and UTZs. Dynamics and mixing cannot be directly explained by this framework due to the absence of rotation and the vertical velocity component in the zone definition. Further, the generic definition of the uniform zones does not capture the rich detail of momentum streaks \citep{Dennis2011}, roller modes \citep{Jimenez2018}, hairpin vortices \citep{Adrian2000}, and temperature fronts \citep{Sullivan2016}, among other features. However, the persistent organization of these complex features consistently produces a stairstep-shaped signature in instantaneous profiles that is well described by UMZs and UTZs as seen in Figs. \ref{fig:umz_detection}d and \ref{fig:utz_detection}d. The critical advantage of the methodology is therefore to quantify the properties of the turbulent eddy organization and relate these properties to mean statistics that result from an ensemble of the instantaneous events.

For instance, the average differences $\Delta U$ and $\Delta \theta$ represent the sharp change in velocity and temperature, respectively, across the thin gradient layers associated with the zone edges. Both $\Delta U$ and $\Delta \theta$ remain proportional to the respective surface flux parameters $u_*$ and $\theta_*$ for the range of stability simulated (Fig. \ref{fig:jump_profiles}). The vertical thickness $H_{u,\theta}$ of the UMZs and UTZs represents either the size of the well-mixed regions or the spacing between the gradient layers. The thickness is proportional to $z$ for the neutral case, consistent with previous experimental observations \citep{Heisel2020}. The zones become progressively thinner and lose their height dependence as the stability regime approaches z-less stratification (Fig. \ref{fig:thickness_profiles}). The scaling behavior in the zone properties indicate that departure from the log law mean profiles are almost entirely due to changes in the eddy size, such that the primary purpose of the M-O similarity relations $\phi_M$ and $\phi_H$ is to account for the reduced zone thickness (Figs. \ref{fig:thickness_profiles}e and \ref{fig:thickness_profiles}f). While the UMZs and UTZs have similar thickness in the surface layer below 0.1$z_i$, i.e. $H_u \approx H_\theta$, the relative eddy intensity $\Delta \theta / \theta_*$ is approximately 20\% smaller than $\Delta U / u_*$, leading to a turbulent Prandtl number below 1 (Fig. \ref{fig:prandtl_number}).

No conclusions can be made regarding the precise magnitudes for either the differences $\Delta U / u_*$ and $\Delta \theta / \theta_*$ in Fig. \ref{fig:jump_profiles} or the thicknesses $H_u / \ell_u$ and $H_\theta / \ell_\theta$ in Fig. \ref{fig:thickness_profiles}. These magnitudes depend on numerous variables discussed in Sec. \ref{subsec:detection} including the resolution, the extent of the local flow volume, the histogram bin width, and the peak detection parameters. Changing these variables leads to fewer or more numerous detected histogram peaks and corresponding changes in the average zone properties. It was found that the changes are proportional across cases if the parameters are consistent, such that the trends in stratification or between velocity and temperature are robust even if the magnitudes are sensitive to the methodology.

It is worth considering the order of magnitude of $\Delta \theta$ representing the temperature fronts, despite the sensitivities for the precise values. The average vertical gradient of temperature across the front in Fig. \ref{fig:interface_profiles}d increases from approximately 0.06 K\,m$^{-1}$ for $z_i/L=$ 1.7 to 0.2 K\,m$^{-1}$ for $z_i/L=$ 6. Both values are significantly larger than the capping inversion (0.01 K\,m$^{-1}$). In contrast, the primary vertical motions associated with roller modes, bursting events, and downward sweeps are proportional to $u_*$, which decreases with increasing stratification. Future research into the dynamics of these features may reveal that the vertical motions are relatively unrestricted within the uniform zones but are unable to penetrate the large density gradients of the local fronts under strong stability, leading to reduced vertical mixing by turbulence and enhanced layering in flow realizations. The instantaneous flow structure and strong local thermal inversions play a central role in this phenomenological picture, which differs from the traditional view that eddies in the stable PBL lose kinetic energy by acting against the relatively weaker mean temperature gradient.

Findings related to the uniform zone size are specific to the vertical thickness, as the horizontal extent of eddies is not quantified here. Figure \ref{fig:xy_fields} provides visual evidence that the turbulent fluctuations lose coherence in the along-wind direction as stratification increases, i.e. they become less ``streaky''. However, coherence in the fluctuations is retained at the scale of the local flow volume $\mathcal{L}_{x^\prime,y}$ used for the zone detection such that the uniform zone signature is present for each case. For neutral boundary layers, multiple UMZs and associated vortical features are often aligned along the extent of the largest motions \citep[see, e.g.,][]{Adrian2000,Marusic2001,Hwang2018}. Figure \ref{fig:xy_fields} suggests that stratification disrupts the largest motions and the alignment of UMZs which leads to a deficit in turbulent energy at low wavenumbers \citep{Kaimal1972}, but the individual zones corresponding to the production scales remain present, at least for the range of stability simulated. Further investigation is required to corroborate these observations.

The present analysis of UMZs and UTZs provides a more concrete description for the organization of turbulent eddies in the stable PBL and how these eddies are related to mean flow properties. It is possible that a similar organization of relatively uniform regions and thin gradient layers is present for more complex flow conditions such as in the presence of wave effects, in the roughness layer of canopies, or around complex terrain, to name a few. However, the persistent orientation of these features into distinct layers may be a specific property of shear-driven and stratified turbulent flows such that it is not practical to extend the same detection methodology to these complex scenarios.

\begin{acknowledgements}
M. H. is financially supported by the US National Science Foundation (NSF-AGS-2031312). The authors gratefully acknowledge the National Center for Atmospheric Research Computational Information Systems Laboratory for computing resources related to the original simulations (http://n2t.net/ark:/85065/d7wd3xhc) and post-processing analysis (doi:10.5065/D6RX99HX).
\end{acknowledgements}

\bibliographystyle{spbasic_updated}     
\bibliography{references} 

\end{document}